\documentclass[prb, twocolumn, superscriptaddress, footinbib, amsmath, amsfonts, amssymb, amsthm]{revtex4-2}

\usepackage[utf8]{inputenc}
\usepackage[T1]{fontenc}
\usepackage[main=english]{babel}

\usepackage{mleftright}
\usepackage{graphicx}
\usepackage{siunitx}
\DeclareSIUnit\angstrom{\text{Å}}
\usepackage[version=4]{mhchem}

\usepackage{bm}
\usepackage{upgreek}
\usepackage{mathrsfs}

\usepackage{hyperref}
\hypersetup{colorlinks=true, linkcolor=blue, citecolor=blue, urlcolor=blue, filecolor=red, pdfstartview=}

\makeatletter
\newcommand*{\defeq}{\mathrel{\rlap{%
\raisebox{0.3ex}{$\m@th\cdot$}}%
\raisebox{-0.3ex}{$\m@th\cdot$}}=}
\makeatother

\newcommand*{\iu}{\mathrm{i}}
\newcommand*{\Elr}{\mathrm{e}}
\newcommand*{\Pauli}{\upsigma}
\newcommand*{\LCs}{\upepsilon}
\newcommand*{\Kd}{\updelta}
\DeclareMathOperator{\Dd}{\updelta}
\newcommand{\one}{\text{\usefont{U}{bbold}{m}{n}1}}
\MakeRobust{\one}
\newcommand*{\hc}{\mathrm{H.c.}}
\newcommand*{\Z}{\mathbb{Z}}

\DeclareMathOperator{\tr}{tr}
\DeclareMathOperator{\Tr}{Tr}
\DeclareMathOperator{\sgn}{sgn}
\DeclareMathOperator{\diag}{diag}
\DeclareMathOperator{\bigO}{\mathcal{O}}
\DeclareMathOperator{\arcctg}{arcctg}

\newcommand*{\abs}[1]{\mleft\lvert {#1} \mright\rvert}
\newcommand*{\dd}[2][]{\mathop{}\!\mathrm{d}^{#1} {#2}}
\newcommand*{\dv}[2]{\frac{\mathrm{d} #1}{\mathrm{d} #2}}
\newcommand*{\pdv}[2]{\frac{\partial #1}{\partial #2}}
\newcommand*{\pdvc}[3]{\mleft.\frac{\partial #1}{\partial #2}\mright|_{#3}}
\newcommand*{\var}[2][]{\mathop{}\!\delta_{#1} {#2}}

\newcommand*{\vdot}{\bm{\cdot}}
\newcommand*{\vcross}{\bm{\times}}
\newcommand*{\grad}{\bm{\nabla}}
\newcommand*{\vb}[1]{\bm{#1}}
\newcommand*{\vu}[1]{\bm{\hat{#1}}}

\newcommand*{\ket}[1]{\mleft\lvert {#1} \mright\rangle}
\newcommand*{\braket}[2]{\mleft\langle {#1} \middle| {#2} \mright\rangle}
\newcommand*{\mel}[3]{\mleft\langle {#1} \middle| {#2} \middle| {#3} \mright\rangle}
\newcommand*{\ev}[1]{\mleft\langle {#1} \mright\rangle}

\DeclareMathOperator{\Ugp}{U}
\DeclareMathOperator{\SU}{SU}

\DeclareMathOperator{\SO}{SO}

\begin{document}
\title{Unconventional superconductivity from electronic dipole fluctuations}
\author{Grgur Palle}
\email{grgur.palle@kit.edu}
\affiliation{Institute for Theoretical Condensed Matter Physics, Karlsruhe Institute of Technology, 76131 Karlsruhe, Germany}
\author{Jörg Schmalian}
\email{joerg.schmalian@kit.edu}
\affiliation{Institute for Theoretical Condensed Matter Physics, Karlsruhe Institute of Technology, 76131 Karlsruhe, Germany}
\affiliation{Institute for Quantum Materials and Technologies, Karlsruhe Institute of Technology, 76131 Karlsruhe, Germany}
\date{\today}
\begin{abstract}
We study electron-electron Coulomb interactions in electronic systems whose Fermi surfaces possess a finite electric dipole density.
Although there is no net dipole moment, we show that electric monopole-dipole interactions can become sufficiently strong in quasi-2D Dirac metals with spin-orbit coupling to induce unconventional odd-parity superconductivity, similar to the Balian-Werthamer state of \ce{^3He-B}.
Hence materials with spin-orbit-induced band inversion, such as the doped topological insulators \ce{Bi2Se3}, \ce{Bi2Te3}, and \ce{SnTe}, are natural candidate materials where our theory could be relevant.
We discuss the conditions for an electric dipole density to appear on the Fermi surface and develop the formalism to describe its coupling to the plasmon field which mediates the Coulomb interaction.
A mechanism for the enhancement of dipolar coupling is then provided for quasi-2D Dirac systems.
Within a large-$N$ renormalization group treatment, we show that the out-of-plane ($z$-axis) dipole coupling is marginally relevant, in contrast to the monopole coupling which is marginally irrelevant.
For physically realistic parameters, we find that dipole fluctuations can get sufficiently enhanced to result in Cooper pairing.
In addition, we establish that the proposed pairing glue is directly measurable in the $z$-axis optical conductivity.
\end{abstract}

\maketitle

The fluctuations of electric dipole moments of electrons are fundamental to understanding a wide variety of systems, ranging from atomic gases and molecules interacting through van der Waals interactions~\cite{Margenau1939, Israelachvili1974, Langbein1974, Parsegian2005, Kaplan2006, Stone2013, Hermann2017}, to small metallic clusters and their cohesive energies~\cite{Garcia1991}, up to solids with sizable contributions to the binding energy and optical conductivity coming from interband dipole excitations~\cite{Hermann2017, Andersson1998, Dion2004, Grimme2011, Klimes2011, Klimes2012, Berland2015}.
From a microscopic point of view, all these effects are due to processes involving electromagnetic interactions among virtual or real excitations that have electric dipole moments.
The above examples usually involve high-energy processes, at least when compared to typical energy scales of collective modes in correlated electron materials.
For electrons near the Fermi level, on the other hand, the Coulomb interactions among them are crucial to facilitating phenomena such as Mott insulation~\cite{Mott1949, Mott1982}, itinerant magnetism~\cite{Moriya1979, Shimizu1981}, and unconventional superconductivity~\cite{Maiti2013}.
This raises two questions.
First, can one sensibly generalize the concept of electronic dipole excitations to states residing on or near the Fermi surface?
And second, can their Coulomb interactions give rise to non-trivial electronic phases, such as superconductivity?

\begin{figure}[t]
\includegraphics[width=\columnwidth]{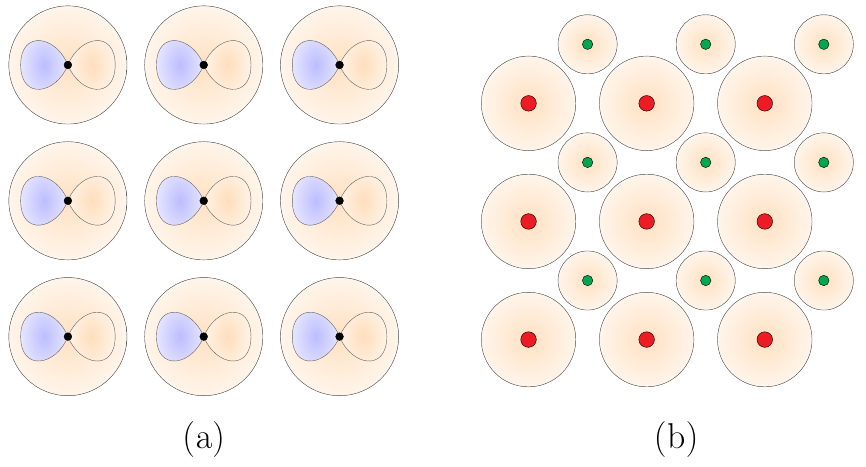} 
\caption{The simplest examples of periodic systems in which local electric dipole operators can be introduced.
This is made possible by the opposite parities of the $s$ and $p_x$ orbitals under (a), and by the different inversions centers (non-trivial Wyckoff positions) of the two $s$ orbitals under (b).}
\label{fig:example-lattices}
\end{figure}

In this paper, we address both of these questions.
We develop the theory of dipole excitations of electronic states near the Fermi surface (Sec.~\ref{sec:dipole-theory}) and we use it to show that the dipolar parts of the Coulomb interaction can result in unconventional superconductivity (Sec.~\ref{sec:pairing}).
In addition, we study Dirac metals (Secs.~\ref{sec:Dirac},~\ref{sec:Dirac-pairing}) as quintessential systems with the two key ingredients for strong Fermi-level dipole effects: parity-mixing, but also strong spin-orbit coupling (SOC), as we explain next.

Electric dipole excitations, while present in generic solids, only contribute to the Fermi surfaces of itinerant systems in the presence of SOC.
To elucidate this important fact, consider a simple lattice with orbitals of opposite parities on each site, such as the $s$ and $p_x$ orbitals shown in Fig.~\ref{fig:example-lattices}(a).
Then in the basis of these two orbitals, a local electric dipole operator $D_x = \uptau_x \otimes \Pauli_0$ exists and is perfectly well-defined.
($\uptau_{\mu}$ and $\Pauli_{\nu}$ are Pauli matrices in orbital and spin space, respectively.)
However, what matters for the description of the itinerant periodic solids is the matrix element
\begin{equation}
\mleft[D_{x; \vb{k} n}\mright]_{ss'} = \mel{u_{\vb{k} n s}}{D_x}{u_{\vb{k} n s'}} \label{eq:dipole-intro}
\end{equation}
in the basis of the Bloch states $u_{\vb{k} n s}$.
Here $\vb{k}$, $n$, and $s$ stand for the crystal momentum, band, and spin, respectively.
In the absence of SOC, the dipole operator is trivial in spin space: $D_{x; \vb{k} n} \propto \Pauli_0$.
It then follows that $D_{x; \vb{k} n} = - D_{x; \vb{k} n} = 0$ for systems invariant under the product $P \Theta$ of space and time inversion.
The same applies to dipole operators constructed in any other way, such as by mixing orbitals of the same parity located at different positions, like in Fig.~\ref{fig:example-lattices}(b).
As we will prove in Sec.~\ref{sec:itinerant-dipoles}, as long as there is no SOC, electric dipole operators vanish when projected onto the Bloch states.
In contrast, with SOC the Fermi surface may acquire a sizable electric dipole density (Fig.~\ref{fig:FS-dip-density}).

The description of electric dipole moments of insulating periodic solids in terms of Bloch states and their Berry connection played an important role in resolving the ambiguity in the definition of the polarization~\cite{KingSmith1993, Resta1992, Resta1993, Vanderbilt1993, Resta1994, Resta2000}.
This description is, in fact, closely related to our treatment of electric dipoles.
As we explain in Sec.~\ref{sec:modern-pol-teo}, the finite extent of the electronic wavefunctions used as a tight-binding basis modifies the periodicity conditions relating $\vb{k} + \vb{G}$ to $\vb{k}$ for inverse lattice vectors $\vb{G}$.
As a result, within the tight-binding basis, the dipole operator as given by the King-Smith--Vanderbilt formula~\cite{KingSmith1993} acquires an anomalous (or intrinsic) contribution
\begin{equation}  
\iu \grad_{\vb{k}} \longrightarrow \iu \grad_{\vb{k}} + \vb{\Gamma}
\end{equation}
which is determined by the same dipole matrix elements that are key to our treatment.
For quasi-2D materials in particular, the anomalous contribution can easily be the dominant one along the out-of-plane direction.

Materials featuring strong SOC and conduction bands which mix parities are therefore natural applications of our theory.
In many materials, such as the topological insulators \ce{Bi2Se3}, \ce{Bi2Te3}, \ce{Sb2Te3}, and \ce{(PbSe)_5(Bi2Se3)_6}~\cite{Ando2013} or the topological crystalline insulators \ce{SnTe} and \ce{Pb_{1-x}Sn_{x}Te}~\cite{Ando2013, Ando2015}, the parity-mixing and SOC come together through SOC-induced band inversion.
As we establish in Sec.~\ref{sec:Dirac-model}, in the vicinity of such band-inverted points, the band structure has essentially the form of a massive Dirac model.
This motivates the investigation of dipole excitations in Dirac metals that we carry out in Sec.~\ref{sec:Dirac}.
Using a large-$N$ renormalization group (RG) analysis of the Coulomb interaction (Sec.~\ref{sec:Dirac-RG}), we show that for quasi-2D Dirac systems, where the monopole coupling is known to be marginally irrelevant~\cite{DTSon2007, Kotov2012}, the $z$-axis dipole coupling becomes marginally relevant.
In Sec.~\ref{sec:Dirac-opticond} we also demonstrate that these enhanced dipole excitations are directly observable in the $z$-axis optical conductivity.

Interestingly, all the materials listed in the previous paragraph become superconductors at low temperatures when doped or pressured~\cite{Note1}.
In the case of doped \ce{Bi2Se3}, there is strong evidence that its superconductivity spontaneously breaks rotational symmetry~\cite{Yonezawa2019, Matano2016, Pan2016, Yonezawa2017, Asaba2017, Du2017, Smylie2018, Cho2020} and has nodal excitations~\cite{Yonezawa2017, Smylie2016, Smylie2017}, indicating an unconventional odd-parity state~\cite{Fu2010, Venderbos2016, Hecker2018}.
Conversely, experiments performed on \ce{In}-doped \ce{SnTe} point towards a fully gapped pairing~\cite{Novak2013, Zhong2013, Smylie2018-p2, Smylie2020} which preserves time-reversal symmetry~\cite{Saghir2014} and has a pronounced drop in the Knight shift~\cite{Maeda2017}.
Although most simply interpreted as conventional $s$-wave pairing, given the moderate change in the Knight shift, a fully-gapped odd-parity state of $A_{1u}$ symmetry is also consistent with these findings~\cite{Smylie2020}.
Because of their topological band structures, these two materials are prominent candidates for topological superconductivity~\cite{Sato2017, Mandal2023}.

\footnotetext[1]{% cite as Note1
Superconductivity under pressure was found in \ce{Bi2Se3}~\cite{Kong2013, Kirshenbaum2013}, \ce{Bi2Te3}~\cite{Zhang2011}, and \ce{Sb2Te3}~\cite{Zhu2013}.
Under ambient pressure, superconductivity was observed in the following compounds doped via intercalation: \ce{Cu_{x}Bi2Se3}~\cite{Hor2010, Kriener2011, Kriener2011-p2}, \ce{Sr_{x}Bi2Se3}~\cite{Liu2015, Shruti2015}, \ce{Nb_{x}Bi2Se3}~\cite{Qiu2015}, \ce{Pd_{x}Bi2Te3} ~\cite{Hor2011}, and \ce{Cu_{x}(PbSe)_5(Bi2Se3)_6}~\cite{Sasaki2014}.
Non-intercalated doping was found to give superconductivity in \ce{Tl_{x}Bi2Se3}~\cite{Wang2016, Trang2016}, \ce{Sn_{1-x}In_{x}Te}~\cite{Sasaki2012, Sato2013}, and \ce{(Pb_{0.5}Sn_{0.5})_{1-x}In_{x}Te}~\cite{Zhong2014, Du2015}.
}

When electric dipole fluctuations are present on the Fermi surface, their monopole-dipole and dipole-dipole interactions can give rise to superconductivity, as we will show in Sec.~\ref{sec:pairing}.
The resulting pairing is necessarily unconventional.
It also requires substantial screening, which is true of most other pairing mechanisms.
Although we find that the dimensionless coupling constant $\lambda$ of the leading pairing channel is comparatively small and not expected to exceed $\sim 0.1$, dipole fluctuations can still be the dominant source of pairing for systems without strong local electronic correlations.
In the case of quasi-2D Dirac metals (Sec.~\ref{sec:Dirac-pairing}), the leading pairing state is an odd-parity state of pseudoscalar ($A_{1u}$) symmetry, similar to the Balian-Werthamer state of \ce{^3He-B}~\cite{Leggett1975, Vollhardt1990, Volovik2003}, while the subleading instability is a two-component $p$-wave state, as required for nematic superconductivity.
Though the latter is the second dominant pairing channel in most cases, it could prevail if aided by a complementary pairing mechanism, such as a phononic one~\cite{Brydon2014, Wu2017}.

Our theory of dipole excitations of Fermi-surface states resembles theories of ferroelectric metals where itinerant electrons couple to ferroelectric modes~\cite{Edge2015, Benedek2016, Chandra2017, KoziiBiRuhman2019, Volkov2020, Enderlein2020, Kozii2022, Volkov2022, Klein2023}, which are usually soft polar phonons~\cite{Cochran1960}.
In both cases, the electrons couple through a fermionic dipole bilinear that is odd under parity and even under time-reversal.
Hence this coupling is direct only in the presence of SOC~\cite{KoziiBiRuhman2019, Volkov2020}, as we already remarked.
However, in our case there is no independent collective mode associated with this dipole bilinear.
Instead, as we show in Sec.~\ref{sec:plasmon-field}, the dipole bilinear contributes to the total charge density alongside a monopole bilinear, and its fluctuations are mediated by the same plasmon field which mediates all electrostatic interactions.
In contrast, ferroelectric modes propagate separately from plasmons and can thus be tuned to quantum criticality, for instance.
As discussed in Ref.~\cite{Klein2023}, this may even give rise to non-Fermi liquid behavior.
We do not expect that such behavior emerges in our theory as dipolar fluctuations will remain massive due to the screening of the Coulomb interaction.

Another distinction between our problem and ferroelectric metals is that, in the Cooper channel, the Coulomb interaction and its monopole-dipole and dipole-dipole parts are repulsive, whereas the exchange of ferroelectric modes is attractive.
The former can therefore only give unconventional pairing (Sec.~\ref{sec:qualitative-pairing}), whereas the latter robustly prefers conventional $s$-wave pairing~\cite{KoziiBiRuhman2019, Enderlein2020, Kozii2022, Volkov2022, Klein2023}, as expected for a type of phonon exchange~\cite{Brydon2014}.
The same distinction applies when comparing our problem to that of metals coupled to more general non-magnetic odd-parity fluctuations~\cite{Kozii2015}.
Apart from this sign difference in the dipole-dipole interaction, a further dissimilarity is that it is the monopole-dipole interaction that is primarily responsible for the pairing in our theory.
This follows from the fact that the dimensionless dipole coupling constant $\tilde{\eta} < 1$ for realistic parameter values so dipole-dipole interactions ($\sim \tilde{\eta}^2$) are weaker than monopole-dipole ones ($\sim \tilde{\eta}$).

In degenerate fermionic gases composed of cold atoms or molecules, electric dipole-dipole interactions have been proposed as a source of pairing in a number of theories~\cite{Baranov2002, Bruun2008, *Bruun2008-E, Sieberer2011, Liu2012, Baranov2012} which appear similar to ours.
Further inspection reveals that they are very different.
A comparison is still instructive.
In these theories, the particles are neutral single-component fermions which carry electric dipole moments.
The electric monopole-dipole interaction, which is key to our mechanism, is thus absent, nor is there any need for screening of the monopole-monopole repulsion.
Their dipole-dipole interaction has no internal structure and its momentum dependence solely determines the preferred pairing channel, whereas in our theory the pseudospin structure of the interaction plays an equally important role.
Their dipoles are also aligned along an external field, giving a net polarization.
In contrast, our electric dipole density varies across the Fermi surface, with opposite momenta and opposite pseudospins having opposite dipole densities (Fig.~\ref{fig:FS-dip-density}).
Finally, unlike in our theory, the nature of their dipole moments is unimportant and one may exchange electric for magnetic dipoles, as has been done experimentally~\cite{Lu2012}.

\section{Theory of dipole excitations of electronic Fermi-surface states} \label{sec:dipole-theory}
Electric dipole moments are conventionally only associated with localized electronic states.
Here, we first show that itinerant electronic states can carry electric dipole moments as well if SOC is present.
After that, we derive the corresponding dipolar contributions to the electron-electron interaction.
Our treatment is then related to the Modern Theory of Polarization.
Lastly, we reformulate the electron-electron Coulomb interaction in terms of a plasmon field, showing that monopole-monopole, monopole-dipole, and dipole-dipole interactions are all mediated by the same plasmon field.

\subsection{Electric dipole moments of itinerant electronic states} \label{sec:itinerant-dipoles}
Itinerant electronic states are states of definite crystal momentum $\vb{k}$, which is defined through the eigenvalues $\Elr^{\iu \vb{k} \vdot \vb{R}}$ of the lattice translation operators $\mathcal{T}_{\vb{R}}$.
Crystal momentum, however, is not the same as physical momentum, the eigenvalue of the continuous translation generator $\vb{P} = - \iu \grad$.
Because of this difference, itinerant electronic states carry not only electric charge and spin, which commute with $\vb{P}$, but also the generalized charges associated with any Hermitian operator $\mathcal{Q}$ that is periodic in the lattice, i.e., that commutes with $\mathcal{T}_{\vb{R}}$.

For instance, the Bloch state 
\begin{equation}
\psi_{\vb{k} n s}(\vb{r}) = \Elr^{\iu \vb{k} \vdot \vb{r}} u_{\vb{k} n s}(\vb{r}) \label{eq:Bloch}
\end{equation}
carries the charge
\begin{equation}
\mel{u_{\vb{k} n s}}{\mathcal{Q}}{u_{\vb{k} n s}} = \int_{\vb{r}} u_{\vb{k} n s}^{\dag}(\vb{r}) Q(\vb{r}) u_{\vb{k} n s}(\vb{r})
\end{equation}
when $\mathcal{Q}(\vb{r}) = \mathcal{N}^{-1} \sum_{\vb{R}} Q(\vb{r}-\vb{R})$; here $\mathcal{N} = \sum_{\vb{R}} 1$ is the number of unit cells and $\int_{\vb{r}} = \int \dd[d]{r}$ goes over all space.
Within tight-binding descriptions, a possible generalized charge is the orbital composition of the Bloch waves.
However, generalized charges associated with electric or magnetic multipoles, local charge or current patterns, and more broadly collective modes in the particle-hole sector of all types are also possible.

\begin{figure}[t]
\includegraphics[width=\columnwidth]{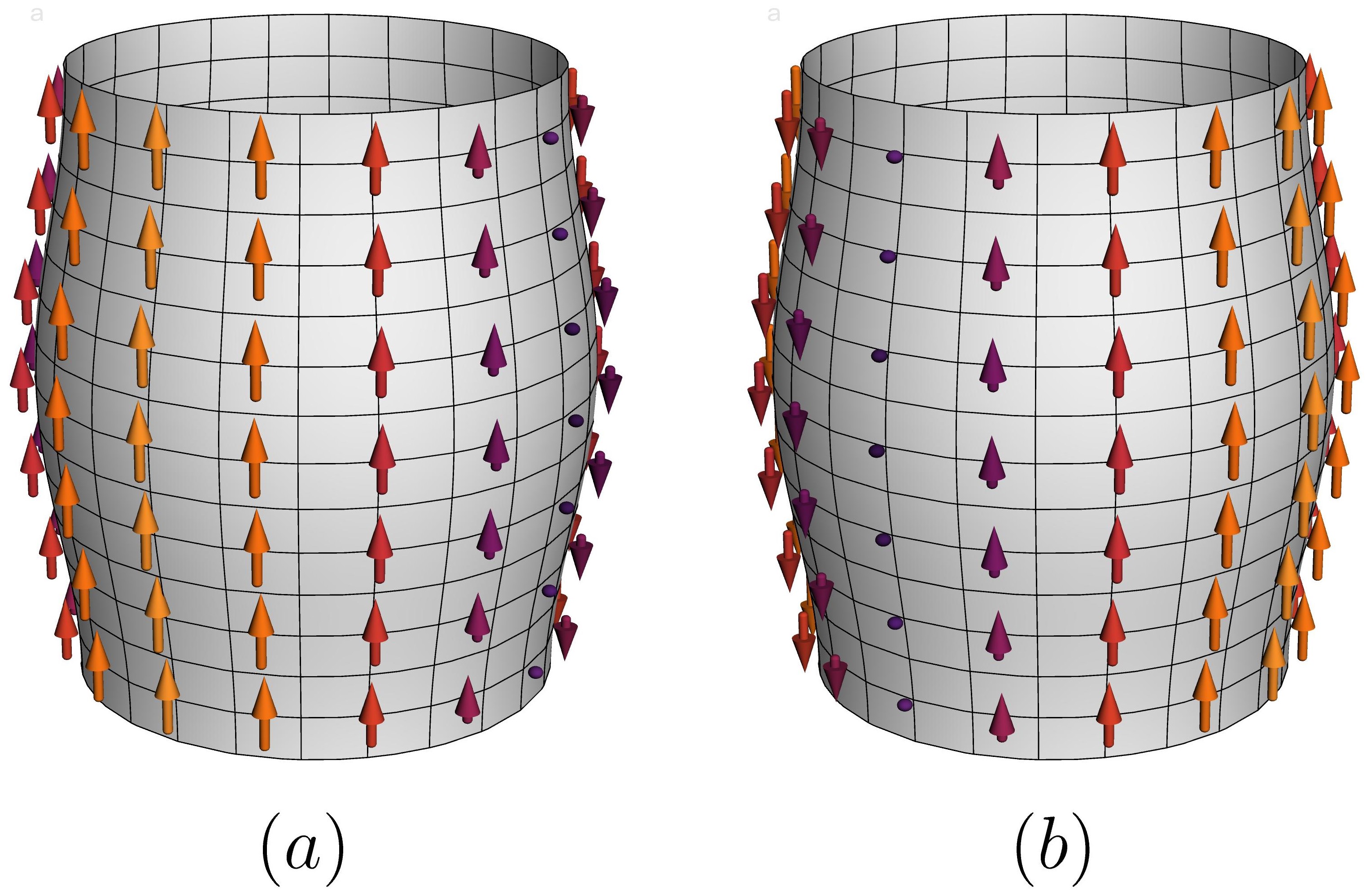}
\caption{An example of a cylindrical Fermi surface with a finite electric dipole density along the $\vu{z}$ direction.
The arrows indicate the direction and strength of the electric dipole density $\mleft[D_{\vu{z}; \vb{k} n}\mright]_{ss}$ for the pseudospin $s$ directed along $+\vu{x}$ (a) and $+\vu{y}$ (b).
Opposite pseudospins and opposite momenta have opposite electric dipole densities.}
\label{fig:FS-dip-density}
\end{figure}

Collective modes couple to their associated generalized charges.
Because they exchange momentum with the electrons, the key matrix elements to analyze are
\begin{equation}
\mel{u_{\vb{k} n s}}{\mathcal{Q}}{u_{\vb{k}+\vb{q} n' s'}}
\end{equation}
of which the dipole element~\eqref{eq:dipole-intro} is a special case with $\vb{q} =  \vb{0}$ and $n' = n$.
At finite $\vb{q}$, or alternatively for $n' \neq n$, these matrix elements are generically finite.
However, the coupling to the Fermi-level electrons ($n' = n$) is particularly strong when they remain finite in the limit $\vb{q} \to \vb{0}$.
This is the limit we discuss in what follows.

In systems without SOC, the periodic parts of the Bloch wavefunctions $\ket{u_{\vb{k} n s}} = \ket{u_{\vb{k} n}} \otimes \ket{s}$ decompose into an orbital and spin part.
Since the composed space and time inversion operation $P \Theta$ is the only symmetry that maps generic $\vb{k}$ to themselves, this is the only symmetry that limits the types of generalized charges that itinerant states can carry.
For a purely orbital charge $\mathcal{Q} = Q \otimes \Pauli_0$ that has sign $p_P \in \{\pm 1\}$ under parity and $p_{\Theta} \in \{\pm 1\}$ under time-reversal, one readily finds the $P \Theta$ symmetry constraint to be
\begin{align}
\mel{u_{\vb{k} n}}{Q}{u_{\vb{k} n}} &= p_P p_{\Theta} \mel{u_{\vb{k} n}}{Q}{u_{\vb{k} n}}.
\end{align}
Hence in the orbital sector only generalized charges with $p_P = p_{\Theta}$ are allowed.
In the spin sector an additional minus sign appears during time inversion so the generalized charges must satisfy $p_P = - p_{\Theta}$ to be finite.
Thus quite generically, a theory of itinerant electronic states that couple without SOC to collective modes as $\vb{q} \to \vb{0}$ is a theory of charge ($p_P = p_{\Theta} = +1$), spin ($p_P = - p_{\Theta} = +1$), and their currents.

Because their $p_P = -1 \neq p_{\Theta} = +1$, electric dipole moments cannot be carried by itinerant electronic states in the absence of SOC (cf.~\cite{KoziiBiRuhman2019, Volkov2020}) and, as a result, they tend to be negligible in most Fermi liquids.
Another recently discussed example are even-parity loop currents ($p_P = +1$, $p_{\Theta} = -1$) which also decouple from electrons in the $\vb{q} \to \vb{0}$ limit~\cite{Palle2024}, consistent with our general scheme.

With spin-orbit coupling, restrictions are much less stringent and generalized charges such as electric dipoles can be carried.
The main difference from the case without SOC is that even-parity orbital operators that commute with the physical spin can acquire a non-trivial structure in pseudospin (Kramers' degeneracy) space.
Conversely, purely spin operators can have trivial pseudospin structures.
In the gauge $\ket{u_{\vb{k} n \downarrow}} = P \Theta \ket{u_{\vb{k} n \uparrow}}$, where $s \in \{\uparrow, \downarrow\}$ are pseudospins, the $P \Theta$ symmetry constraint has the form
\begin{align}
\Pauli_y Q_{\vb{k} n}^{*} \Pauli_y = p_P p_{\Theta} Q_{\vb{k} n},
\end{align}
where $\mleft[Q_{\vb{k} n}\mright]_{ss'} \defeq \mel{u_{\vb{k} n s}}{\mathcal{Q}}{u_{\vb{k} n s'}}$ and $\Pauli_y$ acts in pseudospin space.
Hence $p_P p_{\Theta}$ determines whether $Q_{\vb{k} n}$ is a pseudospin singlet or triplet.
In both cases, $Q_{\vb{k} n}$ can be finite for all $\mathcal{Q}$.

Electric dipoles are pseudospin triplets.
Given their purely orbital nature, this means that SOC need to be relatively strong near the Fermi surface for the electric dipole density to be large.
There is no net electric dipole moment, however.
The total electric dipole density averages to zero at each $\vb{k}$ because of $[D_{\vu{e}; \vb{k} n}]_{\downarrow \downarrow} = - [D_{\vu{e}; \vb{k} n}]_{\uparrow \uparrow}$, where $D_{\vu{e}}$ is the electric dipole operator along the $\vu{e}$ direction.
This is also true for each pseudospin individually in the gauge $\ket{u_{-\vb{k} n s}} = P \ket{u_{\vb{k} n s}}$ since oddness under parity then implies $[D_{\vu{e}; -\vb{k} n}]_{s s} = - [D_{\vu{e}; \vb{k} n}]_{s s}$.
In the simplest case when the point group symmetry matrices are momentum-independent, one finds that $D_{\vu{e}; \vb{k} n} \propto \vu{e} \vdot (\vb{k} \vcross \vb{\Pauli})$~\cite{[{}][{. Note: contrary to what is claimed, the ``Manifestly Covariant Bloch Basis'' does not exist across the whole Brillouin zone in general systems; e.g., if the parity of all time-reversal invariant momenta is not the same.}]{Fu2015}}.
An example of a Fermi surface with an electric dipole density is drawn in Fig.~\ref{fig:FS-dip-density} for the case of a quasi-2D Dirac metal of the type we study in Sec.~\ref{sec:Dirac}.

\subsection{Coulomb interactions and electronic dipole excitations} \label{sec:dipole-coupling}
Here we derive how itinerant electrons which carry electric monopole and dipole moments interact.
Our starting point is the electron-electron Coulomb interaction:
\begin{align}
H_C &= \frac{1}{2} \int_{\vb{r}, \vb{r}'} \rho_e(\vb{r}) \frac{1}{4 \pi \epsilon_0 \abs{\vb{r} - \vb{r}'}} \rho_e(\vb{r}').
\end{align}
The electronic charge density operator is given by
\begin{align}
\rho_e(\vb{r}) = - e \sum_s \Psi_s^{\dag}(\vb{r}) \Psi_s(\vb{r}),
\end{align}
where $s \in \{\uparrow, \downarrow\}$ are the physical spins.

Next, we expand the fermionic field operators in a complete lattice basis:
\begin{align}
\Psi_s(\vb{r}) &= \sum_{\vb{R} \alpha} \mleft[\varphi_{\alpha}(\vb{r} - \vb{R})\mright]_s \psi_{\alpha}(\vb{R}).
\end{align}
Here, we allow the basis to depend on spin and $\alpha$ is a combined orbital and spin index.
One popular choice of basis functions are the Wannier functions~\cite{Marzari2012}.
If they are constructed from a set of bands which (i) has vanishing Chern numbers and (ii) does not touch any of the bands of the rest of the spectrum, then the corresponding Wannier functions can always be made exponentially localized~\cite{Panati2007, Brouder2007}.
Condition (i) is always satisfied in the presence of time-reversal symmetry, while the second condition can be satisfied to an adequate degree by including many bands.
Thus as long as we do not restrict ourselves to the description of low-energy bands, we may assume that our basis functions $\varphi_{\alpha}(\vb{r} - \vb{R})$ are exponentially localized.
Using this basis, we may now decompose the charge density into localized parts:
\begin{align}
\rho_e(\vb{r}) = \sum_{\vb{R}} \rho_{\vb{R}}(\vb{r} - \vb{R}),
\end{align}
where the $\rho_{\vb{R}}(\vb{r})$ are localized around $\vb{r} = \vb{0}$:
\begin{align}
\begin{aligned}
\rho_{\vb{R}}(\vb{r}) &= - \frac{e}{2} \sum_{\vb{\Delta} \alpha_1 \alpha_2} \varphi_{\alpha_1}^{\dag}(\vb{r}) \varphi_{\alpha_2}(\vb{r} - \vb{\Delta}) \\[-6pt]
&\hspace{58pt} \times \psi_{\alpha_1}^{\dag}(\vb{R}) \psi_{\alpha_2}(\vb{R} + \vb{\Delta}) + \hc
\end{aligned}
\end{align}
Here the $\vb{\Delta}$ sum goes over lattice neighbors.

By expanding $H_C$ to dipolar order in multipoles, we obtain
\begin{equation}
H_{\text{int}} = \frac{1}{2} \sum_{\vb{R}_1 \vb{R}_2} \sum_{\mu \nu} D_{\mu}(\vb{R}_1) V_{\mu \nu}(\vb{R}_1 - \vb{R}_2) D_{\nu}(\vb{R}_2), \label{eq:Hint-vR}
\end{equation}
where $\mu, \nu \in \{0, 1, 2, 3\}$,
\begin{equation}
D_0(\vb{R}) = \int_{\vb{r}} \rho_{\vb{R}}(\vb{r})
\end{equation}
is the electric monopole moment operator, and 
\begin{equation}
D_i(\vb{R}) = \int_{\vb{r}} r_i \rho_{\vb{R}}(\vb{r})  
\end{equation}
are the components of the electric dipole operator.
The integration $\int_{\vb{r}} = \int \dd[3]{r}$ extends over the whole space, not over a unit cell.
Due to exponential localization, these integrals converge and give well-defined operators.
Because we are working with a non-periodic $\rho_{\vb{R}}(\vb{r})$, there is no ambiguity in these definitions, other than the obvious dependence on the choice of basis functions $\varphi_{\alpha}(\vb{r} - \vb{R})$.

The interaction matrix which follows from the multipole expansion equals
\begin{equation}
V_{\mu \nu}(\vb{R}) = \frac{1}{4 \pi \epsilon_0} \begin{pmatrix}
1 & - \partial_j \\
\partial_i & - \partial_i \partial_j
\end{pmatrix} \frac{1}{R}.
\end{equation}
Here, $R = \abs{\vb{R}}$, $i, j \in \{1, 2, 3\}$, and $\partial_i = \partial/\partial{R_i}$. 
At $\vb{R} = \vb{0}$, $V_{\mu \nu}(\vb{R})$ has an aphysical divergence that we regularize by replacing $R^{-1}$ with $R^{-1} \operatorname{erf}\frac{R}{2 r_0}$; this corresponds to an unscreened Hubbard interaction $U = e^2/(4 \pi^{3/2} \epsilon_0 r_0)$.
The Fourier transform $q^{-2} \Elr^{- r_0^2 q^2}$ then decays exponentially for large $q = \abs{\vb{q}}$, making the Umklapp sum, that occurs upon Fourier transforming $V_{\mu \nu}(\vb{R})$, convergent.
For $r_0$ small compared to the lattice constant, the Umklapp sum is well-approximated with just the $\vb{G} = \vb{0}$ term.
Hence in momentum space:
\begin{align}
H_{\text{int}} &= \frac{1}{2 L^d} \sum_{\vb{q} \mu \nu} D_{\mu}(-\vb{q}) V_{\mu \nu}(\vb{q}) D_{\nu}(\vb{q}), \label{eq:Hint-vq}
\end{align}
where $L^d$ is the total volume in $d$ spatial dimensions, $\vb{q}$ goes over the first Brillouin zone, and
\begin{align}
V_{\mu \nu}(\vb{q}) &= \begin{pmatrix}
1 & - \iu q_j \\
\iu q_i & q_i q_j
\end{pmatrix} \frac{1}{\epsilon_0 \vb{q}^2}. \label{eq:VUmklapp}
\end{align}
Keeping only the $\vb{G} = \vb{0}$ Umklapp term in $V_{\mu \nu}(\vb{q})$ can also be understood as another way of regularizing the $V_{\mu \nu}(\vb{R} = \vb{0})$ divergence.
When we later consider quasi-2D systems, the Umklapp sum for the out-of-plane $\vb{G}$ will not be negligible.
Its main effect is to make $V_{\mu \nu}(\vb{q})$ periodic in the out-of-plane $q_z$, which we shall later account for by replacing all $q_z$ with $\sin q_z$.

For the $D_{\mu}(\vb{R})$, we now obtain, in matrix notation,
\begin{align}
D_{\mu}(\vb{R}) &= - \frac{e}{2} \psi^{\dag}(\vb{R}) \Gamma_{\mu}(\vb{\Delta}) \psi(\vb{R} + \vb{\Delta}) + \hc, \label{eq:dipole_real_space}
\end{align}
where
\begin{align}
\mleft[\Gamma_{0}(\vb{\Delta})\mright]_{\alpha_1 \alpha_2} &\defeq \int_{\vb{r}} \varphi_{\alpha_1}^{\dag}(\vb{r}) \varphi_{\alpha_2}(\vb{r} - \vb{\Delta}), \\
\mleft[\Gamma_{i}(\vb{\Delta})\mright]_{\alpha_1 \alpha_2} &\defeq \int_{\vb{r}} r_i \varphi_{\alpha_1}^{\dag}(\vb{r}) \varphi_{\alpha_2}(\vb{r} - \vb{\Delta}). \label{eq:dipol-matrix-elem}
\end{align}
When the lattice bases $\varphi_{\alpha}(\vb{r} - \vb{R})$ are orthogonal and normalized $\Gamma_{0}(\vb{\Delta}) = \Kd_{\vb{\Delta}, \vb{0}} \one$, and when they are sufficiently localized $\Gamma_{i}(\vb{\Delta}) \approx 0$ for $\vb{\Delta}$ which are not $\vb{0}$ or the nearest-lattice neighbors.
Moreover, $[\Gamma_{i}(\vb{\Delta}=\vb{0})]_{\alpha_1 \alpha_2}$ is finite for $\varphi_{\alpha_{1,2}}(\vb{r})$ centered at $\vb{r} = \vb{0}$ only when they have opposite parities.
That said, substantial dipole moments can also arise from orbitals of the same parity if they belong to different neighboring atoms because of the possibility of forming anti-binding superpositions; see Fig.~\ref{fig:example-lattices}(b).

In the simplest case when only $\Gamma_{0}(\vb{\Delta}=\vb{0}) \equiv \Gamma_0 = \one$ and $\Gamma_{i}(\vb{\Delta}=\vb{0}) \equiv \Gamma_i$ are finite, in momentum space we get
\begin{align}
D_{\mu}(\vb{q}) &= - e \sum_{\vb{k}} \psi^{\dag}(\vb{k}) \Gamma_{\mu} \psi(\vb{k} + \vb{q}),
\end{align}
where $\vb{k}$ runs over the first Brillouin zone.
The associated matrix elements $\mel{u_{\vb{k} n s}}{\Gamma_{\mu}}{u_{\vb{k} + \vb{q} n' s'}}$ were analyzed in the previous section.
The monopole matrix elements ($\mu = 0$) become diagonal in the band index as $\vb{q} \to \vb{0}$, but are otherwise finite.
The intraband dipole matrix elements ($\mu = 1, 2, 3$), on the other hand, vanish in the $\vb{q} \to \vb{0}$ limit in the absence of SOC.
The corresponding coupling of the electric dipoles to Fermi-level electrons thus gains an additional momentum power, which makes these interactions even more irrelevant with respect to RG flow than usual, unless the system has spin-orbit coupling.

The multipole expansion employed in Eq.~\eqref{eq:Hint-vR} is justified whenever two charges are localized on length scales smaller than their distance.
In the limit of strong screening that we later analyze, however, the strongest interactions come from nearby particles, indicating a breakdown of the multipole expansion.
Nonetheless, the additional dipolar terms that we identified in the effective electron-electron interaction of Eq.~\eqref{eq:Hint-vR} will still be present, albeit with coefficients that are phenomenological parameters.
Although their values cannot be inferred from a direct multipole expansion when screening is strong, the momentum-dependence and form of the dipolar coupling follows from symmetry and retains the same structure as derived in this section.

\subsection{Relation to the Modern Theory of Polarization} \label{sec:modern-pol-teo}
Our theory deals with dynamical electric dipole moments and their fluctuations.
Nonetheless, it is enlightening to make contact to the Modern Theory of Polarization~\cite{KingSmith1993, Resta1992, Resta1993, Vanderbilt1993, Resta1994, Resta2000, Nissinen2021} in which the static polarization is expressed in terms of the Berry connection via~\cite{KingSmith1993}
\begin{equation}
\ev{\vb{D}} = - e \sum_{\vb{k} n s}^{\text{occ.}} \mel{u_{\vb{k} n s}}{\iu \grad_{\vb{k}}}{u_{\vb{k} n s}}, \label{eq:KingSmith}
\end{equation}
where $\vb{k}$ goes over the first Brillouin zone, $n$ goes over occupied bands only, and $s \in \{\uparrow, \downarrow\}$ is the pseudospin.
As we show below, the finite extent of the $\varphi_{\alpha}(\vb{r} - \vb{R})$ basis wavefunctions, which is crucial for the definition of the higher-order multipoles in the first place, gives rise to an anomalous contribution to the polarization when expressed within a tight-binding description.

Assuming time-reversal symmetry, the Bloch wavefunctions of Eq.~\eqref{eq:Bloch} can always be chosen to be periodic in $\vb{k}$~\cite{Note2}, meaning $\psi_{\vb{k} + \vb{G} n s}(\vb{r}) = \psi_{\vb{k} n s}(\vb{r})$ for all inverse lattice vectors $\vb{G}$, where $\psi_{\vb{k} n s}(\vb{r})$ are continuous, but not necessarily analytic, functions of $\vb{k}$.
The real-space periodic parts $u_{\vb{k} n s}(\vb{r}) = u_{\vb{k} n s}(\vb{r} + \vb{R})$ then satisfy
\begin{equation}
u_{\vb{k} n s}(\vb{r}) = \Elr^{\iu \vb{G} \vdot \vb{r}} u_{\vb{k} + \vb{G} n s}(\vb{r}). \label{eq:k-period-cond}
\end{equation}
Next, we expand the $u_{\vb{k} n s}(\vb{r})$ with respect to an orthonormal tight-binding basis:
\begin{equation}
u_{\vb{k} n s}(\vb{r}) = \sum_{\vb{R} \alpha} \varphi_{\alpha}(\vb{r} - \vb{R}) \mleft[v_{\vb{k} n s}\mright]_{\alpha}.
\end{equation}
The periodicity condition~\eqref{eq:k-period-cond} now becomes:
\begin{equation}
v_{\vb{k} n s} = U_{\vb{G}} \, v_{\vb{k} + \vb{G} n s}, \label{eq:k-period-cond2}
\end{equation}
where
\begin{equation}
\mleft[U_{\vb{G}}\mright]_{\alpha_1 \alpha_2} = \sum_{\vb{\Delta}} \int_{\vb{r}} \varphi_{\alpha_1}^{\dag}(\vb{r}) \Elr^{\iu \vb{G} \vdot \vb{r}} \varphi_{\alpha_2}(\vb{r} - \vb{\Delta}).
\end{equation}
In evaluating this expression, one often argues that the wavefunctions are point-like objects such that $\Elr^{\iu \vb{G} \vdot \vb{r}} \varphi_{\alpha}(\vb{r}) \approx \Elr^{\iu \vb{G} \vdot \vb{x}_{\alpha}} \varphi_{\alpha}(\vb{r})$, where $\vb{x}_{\alpha}$ are the positions of the orbitals; see also Refs.~\cite{Fruchart2014, Simon2020}.
This would then give a diagonal $\mleft[U_{\vb{G}}\mright]_{\alpha_1 \alpha_2} = \Elr^{\iu \vb{G} \vdot \vb{x}_{\alpha_1}} \Kd_{\alpha_1 \alpha_2}$ with $\Ugp(1)$ phase factors which can be absorbed into the $\mleft[v_{\vb{k} n s}\mright]_{\alpha}$ through a $\Ugp(1)$ gauge transformation.
However, the spread of the $\varphi_{\alpha}(\vb{r})$ around $\vb{x}_{\alpha}$ also contributes significantly to $U_{\vb{G}}$ when the orbitals mix parities or overlap.
By expanding the $\Elr^{\iu \vb{G} \vdot \vb{r}}$ exponential to linear order in $\vb{r}$, one readily finds that these corrections result in
\begin{equation}
U_{\vb{G}} = \Elr^{\iu \vb{G} \vdot \sum_{\vb{\Delta}} \vb{\Gamma}(\vb{\Delta})}, \label{eq:U-G-expression}
\end{equation}
where the $\Gamma_{i}(\vb{\Delta})$ are the matrix elements of Eq.~\eqref{eq:dipol-matrix-elem}.
Having found tight-binding vectors $v_{\vb{k} n s}^{(0)}$ that are periodic, $v_{\vb{k} + \vb{G} n s}^{(0)} = v_{\vb{k} n s}^{(0)}$, the periodicity condition~\eqref{eq:k-period-cond2} can be accommodated by the $\Ugp(N)$ transformation
\begin{equation}
v_{\vb{k} n s} = \Elr^{- \iu \vb{k} \vdot \sum_{\vb{\Delta}} \vb{\Gamma}(\vb{\Delta})} v_{\vb{k} n s}^{(0)}.
\end{equation}
This holds to the same order in momentum as the expression for $U_{\vb{G}}$~\cite{Note3}.
Within the $\varphi_{\alpha}(\vb{r} - \vb{R})$ basis, the King-Smith--Vanderbilt formula~\eqref{eq:KingSmith} therefore acquires an additional term:
\begin{equation}
\ev{\vb{D}} = - e \sum_{\vb{k} n s}^{\text{occ.}} \mel{v_{\vb{k} n s}^{(0)}}{\iu \grad_{\vb{k}} + \sum\nolimits_{\vb{\Delta}} \vb{\Gamma}(\vb{\Delta})}{v_{\vb{k} n s}^{(0)}}. \label{eq:anomal-KingSmith}
\end{equation}
This additional, or anomalous, term is determined by the same $\Gamma_{i}(\vb{\Delta})$ of Eq.~\eqref{eq:dipol-matrix-elem} that govern the dipolar interactions.

\footnotetext[2]{% cite as Note2
This follows from the fact that the band energies are bounded from below, thereby precluding spectral flow in which the $n$-th band maps to a different band as loops are traversed in the Brillouin zone.
This remains true even if the bands cross, albeit with a non-analytic $\vb{k}$-dependence around the crossing.
If a (possibly degenerate) band with vanishing Chern numbers does not touch any other band, one can always choose a gauge in which $\psi_{\vb{k} n s}$ and $u_{\vb{k} n s}$ are analytic functions of $\vb{k}$~\cite{Panati2007}.
}

\footnotetext[3]{% cite as Note3
Note that the $\Gamma_{i} = \sum_{\vb{\Delta}} \Gamma_{i}(\vb{\Delta})$ matrices do not commute so $\Elr^{\iu \vb{k} \vdot \vb{\Gamma}} \Elr^{\iu \vb{G} \vdot \vb{\Gamma}} \Elr^{- \iu (\vb{k} + \vb{G}) \vdot \vb{\Gamma}} = \Elr^{\tfrac{1}{2} [\vb{k} \vdot \vb{\Gamma}, \vb{G} \vdot \vb{\Gamma}] + \cdots} \neq \one$.
}

To illustrate the importance of this anomalous term, let us consider a system whose tight-binding Hamiltonian is independent of $k_z$.
This is often approximately true in quasi-2D systems.
The eigenvectors $v_{\vb{k} n s}^{(0)}$ are then independent of $k_z$ and a naive application of Eq.~\eqref{eq:KingSmith} would suggest that the out-of-plane polarization vanishes.
However, Eq.~\eqref{eq:anomal-KingSmith} reveals that this is not necessarily true:
\begin{equation}
\ev{D_z} = - e \sum_{\vb{k} n s}^{\text{occ.}} \sum_{\vb{\Delta}} \mel{v_{\vb{k} n s}^{(0)}}{\Gamma_z(\vb{\Delta})}{v_{\vb{k} n s}^{(0)}}
\end{equation}
can be finite when the wavefunctions are spread along the $\vu{z}$ direction, even though there is no hopping along $z$.
In Dirac systems, this regime, which is dominated by the anomalous term, will turn out to have the strongest enhancement of dipole fluctuations, as we show in Sec.~\ref{sec:Dirac}.

\subsection{Formulation in terms of a plasmon field} \label{sec:plasmon-field}
Here we reformulate the effective interaction $H_{\text{int}}$ of Eq.~\eqref{eq:Hint-vR} in terms of Hubbard-Stratonovich (HS) fields~\cite{Altland2010}.
Naively, one would do this by introducing a HS field for each component of $D_{\mu}$.
The result would then formally look like the models of ferroelectric critical fluctuations coupled to fermions that have been the subject of much recent interest~\cite{KoziiBiRuhman2019, Volkov2020, Enderlein2020, Kozii2022, Klein2023}.
Specifically, there would be a monopole HS field and an independent dipole HS field with the same symmetry and coupling to fermions as ferroelectric modes.
However, this is not correct for our $H_{\text{int}}$ because the same electrostatic fields mediates all electric interactions, whether they are monopole-monopole, monopole-dipole, or dipole-dipole.
Formally, this manifests itself through the non-invertible rank~1 matrix structure of $V_{\mu \nu}(\vb{q})$ in Eq.~\eqref{eq:VUmklapp}.
Within perturbation theory, one may indeed confirm that this rank~1 matrix structure stays preserved and that only $\epsilon_0 \vb{q}^2 \to \epsilon(\vb{q}) \vb{q}^2$ gets renormalized.

\begin{figure}[t]
\includegraphics[width=\columnwidth]{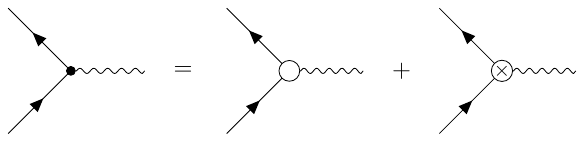}
\caption{Decomposition of the total electron-plasmon vertex (solid dot) into a monopole-plasmon (open circle) and dipole-plasmon (crossed circle) contribution, as follows from the expansion of the electron density of Eq.~\eqref{eq:multipole-exp}.
Solid lines stand for electrons and wiggly lines for plasmons.}
\label{fig:plasmon-fermion-vertex}
\end{figure}

To carry out the HS transformations, we group all $D_{\mu}$ into one effective charge density:
\begin{align}
\rho(\vb{q}) &= D_0(\vb{q}) - \iu \sum_j q_j D_j(\vb{q}). \label{eq:multipole-exp}
\end{align}
If we were not on a lattice, in real space this expression would reduce to the familiar $\rho(\vb{r}) = D_0(\vb{r}) - \grad \vdot \vb{D}(\vb{r})$, with $D_0$ playing the role of the free charge density and $\vb{D}$ the role of the polarization density.
The Euclidean action of $H_{\text{int}}$ is
\begin{align}
\mathcal{S}_{\text{int}}[\psi] &= \frac{1}{2 \beta L^d} \sum_{q} \rho(-q) V(q) \rho(q),
\end{align}
where $q = (\omega_q, \vb{q})$, $\omega_q \in 2 \pi \Z / \beta$ are bosonic Matsubara frequencies, and $V(q) = V_{00}(\vb{q}) = 1 / (\epsilon_0 \vb{q}^2)$.
After the HS transformation it becomes:
\begin{align}
\begin{aligned}
\mathcal{S}_{\text{int}}[\phi, \psi] &= \frac{1}{2} \sum_{q} \phi(-q) V^{-1}(q) \phi(q) \\[-4pt]
&\hspace{14pt} + \frac{\iu}{\sqrt{\beta L^d}} \sum_{q} \phi(-q) \rho(q),
\end{aligned} \label{eq:HS-int-action}
\end{align}
where $\phi(q) = \phi^{*}(-q)$ is the electrostatic (plasmon) field.
The only difference from the usual HS-formulated action of plasma excitations is that the charge density has additional contributions coming from itinerant electric dipoles.
This is illustrated in Fig.~\ref{fig:plasmon-fermion-vertex}, where we show the decomposition of the total electron-plasmon vertex into monopole-plasmon and dipole-plasmon contributions, in agreement with the expansion of Eq.~\eqref{eq:multipole-exp}.

\section{Dipole excitations in Dirac metals} \label{sec:Dirac}
In many systems, the electric dipole moments are relatively small, and if the spin-orbit coupling (SOC) is weak, their contribution to the interaction of Fermi surface states is even smaller.
Yet in Dirac systems which are generated through band inversion the opposite is the case.
Band inversion takes place when SOC inverts bands of opposite parities near high-symmetry points.
This large mixing of parities enables large electric dipole moments which, due to strong SOC, project onto the Fermi surface to significantly modify the electrostatic interaction.
Dirac metals therefore provide fertile ground for sizable electric dipole effects.

Here, we first show that the band Hamiltonian describing the vicinity of band-inverted points has the form of an anisotropic gapped Dirac model.
We then derive how the electric dipole moments are represented within this model (Table~\ref{tab:gamma-class}) and we introduce the corresponding electrostatic interactions of Sec.~\ref{sec:dipole-coupling} to the model.
In Sec.~\ref{sec:Dirac-opticond}, we turn to the study of the polarization of this model in the quasi-2D limit of weak $z$-axis dispersion.
Although it should naively vanish in this limit, we show that the additional dipole coupling renders the $z$-axis optical conductivity finite, thereby opening a route towards experimentally measuring the dipole excitations of our theory.
We then use renormalization group (RG) methods to investigate the dipole-coupled Dirac model in the regime of strong screening (Sec.~\ref{sec:Dirac-RG}, Fig.~\ref{fig:RG-sketch}).
This regime coincides with strong coupling and to access it analytically, we employ a large-$N$ expansion to $1$-loop order, $N$ being the number of fermion flavors.
For generic Fermi surfaces, we find that electric dipole coupling is irrelevant in the RG sense and thus becomes weaker at low energies.
However, if the dipole moments are parallel to the Fermi surface, as is the case for the out-of-plane moments in quasi-2D systems, they are marginal.
A detailed analysis furthermore shows that they are marginally relevant (Fig.~\ref{fig:RG-etaz-flow}), in contrast to the monopole coupling constant which is marginally irrelevant (Fig.~\ref{fig:RG-alpha-flow}).
Note that the dispersion along the out-of-plane direction here needs to be flat on the scale of the band gap of the semimetal because otherwise $z$-axis scaling would tend to make the $z$-axis dipole moments irrelevant.

\subsection{The model} \label{sec:Dirac-model}
The minimal model which captures the essential physics and that we shall study has the Euclidean (imaginary time) action
\begin{align}
\mathcal{S}[\psi, \phi] &= \mathcal{S}_{\psi}[\psi] + \mathcal{S}_{\phi}[\phi] + \mathcal{S}_{\psi\phi}[\psi, \phi], \label{eq:model-action}
\end{align}
where $\mathcal{S}_{\psi}$ and $\mathcal{S}_{\phi}$ are the non-interacting fermionic and plasmonic parts, and $\mathcal{S}_{\psi\phi}$ describes the electrostatic coupling between the two.

To construct the fermionic part, we consider two bands of opposite parities in the vicinity of the $\Gamma$ point $\vb{k} = \vb{0}$.
The parity and time-reversal transformation matrices are
\begin{align}
P &= \uptau_3 \Pauli_0, &
P \Theta &= \uptau_0 \iu \Pauli_y,
\end{align}
where $\uptau_a$ and $\Pauli_b$ are Pauli matrices in band and pseudospin space, respectively.
In Table~\ref{tab:tausigma-class} we classify all the matrices according to their parity and time-reversal signs.
The only two matrices which are even under both parity and time-reversal are $\uptau_3 \Pauli_0$ and $\uptau_0 \Pauli_0$ and they give the band gap and chemical potential displacement in the Hamiltonian, respectively.

Because of the parity-mixing, terms linear in $\vb{k}$ also arise in the Hamiltonian.
They are constructed by combining $\vb{k}$ with three out of the four parity and time-reversal odd matrices $\uptau_1 \Pauli_0$, $\uptau_2 \Pauli_x$, $\uptau_2 \Pauli_y$, and $\uptau_2 \Pauli_z$; which ones depends on the rotational symmetries.
When there is $n$-fold rotation symmetry around the $z$ axis, with $n \geq 3$, that has the form
\begin{align}
C_{n z} &= \uptau_0 \exp\mleft(- \iu \tfrac{2 \pi}{n} \tfrac{1}{2} \Pauli_z\mright), \label{eq:J12-exp}
\end{align}
the pairs $\mleft(\uptau_2 \Pauli_x, \uptau_2 \Pauli_y\mright)$ and $\mleft(\uptau_2 \Pauli_y, - \uptau_2 \Pauli_x\mright)$ transform the same as $(k_x, k_y)$, giving a Rashba-like term in the Hamiltonian.
When there is twofold rotation symmetry around the $x$ axis, its form determines which of these two pairs continues to transform as $(k_x, k_y)$, as well as whether $\uptau_1 \Pauli_0$ or $\uptau_2 \Pauli_z$ transforms the same as $k_z$.
For
\begin{gather}
\begin{gathered}
C_{2 x}' = \uptau_3 (- \iu \Pauli_x), \\
\text{$\mleft(\uptau_2 \Pauli_y, - \uptau_2 \Pauli_x\mright) \sim (k_x, k_y)$ and $\uptau_1 \Pauli_0 \sim k_z$,}
\end{gathered} \label{eq:J31-exp}
\end{gather}
whereas for
\begin{gather}
\begin{gathered}
C_{2 x}' = \uptau_0 (- \iu \Pauli_x), \\
\text{$\mleft(\uptau_2 \Pauli_x, \uptau_2 \Pauli_y\mright) \sim (k_x, k_y)$ and $\uptau_2 \Pauli_z \sim k_z$.}
\end{gathered} \label{eq:J31-exp-v2}
\end{gather}
For concreteness, below we assume the former [Eq.~\eqref{eq:J31-exp}].
The latter choice for $C_{2 x}'$ is related to the former one through the basis change $\mathcal{B}^{\dag} \mleft.C_{2 x}'\mright|_{\text{Eq.\eqref{eq:J31-exp-v2}}} \mathcal{B} = \mleft.C_{2 x}'\mright|_{\text{Eq.\eqref{eq:J31-exp}}}$, where $\mathcal{B} = \diag(1,1,-\iu,\iu)$.
This basis change preserves the other symmetry matrices ($\mathcal{B}^{\dag} P \mathcal{B} = P$, $\mathcal{B}^{\dag} P \Theta \mathcal{B}^{*} = P \Theta$, and $\mathcal{B}^{\dag} C_{n z} \mathcal{B} = C_{n z}$), which implies that all later results are independent of which $C_{2 x}'$ we use.

\begin{table}[t]
\caption{The classification of the $16$ orbital and spin matrices $\uptau_a \Pauli_b$ according to parity $\hat{P}$ and time-reversal $\hat{\Theta}$.
Vector $\vec{\Pauli}$ is a shorthand for $\{\Pauli_i\} = \{\Pauli_x, \Pauli_y, \Pauli_z\}$ and $\hat{\mathcal{K}}$ is the complex conjugation operator.}
{\renewcommand{\arraystretch}{1.3}
\renewcommand{\tabcolsep}{4.0pt}
\begin{tabular}{l|c|c|c|c} \hline\hline
& $\uptau_0 \Pauli_0$, $\uptau_3 \Pauli_0$ & $\uptau_0 \vec{\Pauli}$, $\uptau_3 \vec{\Pauli}$ & $\uptau_2 \Pauli_0$, $\uptau_1 \vec{\Pauli}$ & $\uptau_1 \Pauli_0$, $\uptau_2 \vec{\Pauli}$ \\ \hline
$\hat{P} = \uptau_3 \Pauli_0$ & $+1$ & $+1$ & $-1$ & $-1$ \\
$\hat{\Theta} = \uptau_3 \iu \Pauli_y \hat{\mathcal{K}}$ & $+1$ & $-1$ & $+1$ & $-1$
\\ \hline\hline
\end{tabular}}
\label{tab:tausigma-class}
\end{table}

The effective Hamiltonian near $\vb{k} = \vb{0}$ therefore reads
\begin{align}
\begin{aligned}
H_{\vb{k}} &= m \uptau_3 \Pauli_0 + v \uptau_2 (k_x \Pauli_y - k_y \Pauli_x) \\
&\hspace{20pt} + v_z k_z \uptau_1 \Pauli_0 - \mu \uptau_0 \Pauli_0,
\end{aligned} \label{eq:Dirac-Haml}
\end{align}
with the corresponding action being:
\begin{align}
\mathcal{S}_{\psi}[\psi] &= \sum_k \psi^{\dag}(k) \mleft[- \iu \omega_k + H_{\vb{k}}\mright] \psi(k),
\end{align}
where $k = (\omega_k, \vb{k})$ and $\omega_k \in \pi (2 \Z + 1) / \beta$ are fermionic Matsubara frequencies.
Because the $\vb{k}$-linear terms depend on spin, they need SOC to be large.
At quadratic order in $\vb{k}$, $m$ and $\mu$ gain momentum dependence, as do $v$ and $v_z$ at cubic order in $\vb{k}$. 
This does not affect things qualitatively as long as the $\vb{k}$-linear terms are dominant so we shall not include this higher order $\vb{k}$-dependence in our analysis.
We shall also neglect the $\propto \uptau_2 \Pauli_z$ term which arises at cubic order and breaks the emergent Dirac form.

To recast the action more closely as a Dirac model, introduce the Euclidean gamma matrices
\begin{align}
\begin{aligned}
\gamma_0 &= \uptau_3 \Pauli_0, &&&\qquad
\gamma_1 &= \uptau_1 \Pauli_y, \\
\gamma_2 &= - \uptau_1 \Pauli_x, &&&\qquad
\gamma_3 &= - \uptau_2 \Pauli_0.
\end{aligned}
\end{align}
The Euclidean signature we shall use not only for the gamma matrices, $\{\gamma_{\mu}, \gamma_{\nu}\} = 2 \Kd_{\mu \nu}$, but also for raising, lowering, and contracting the indices of any four-vector.
By switching from $\psi^{\dag}$ to $\bar{\psi} = \psi^{\dag} \gamma_0$, one now readily finds that
\begin{align}
\mathcal{S}_{\psi}[\psi] &= \sum_k \bar{\psi}(k) G^{-1}(k) \psi(k), \label{eq:S-psi}
\end{align}
where
\begin{align}
\begin{aligned}
G^{-1}(k) &= m \one - \iu \mleft[\omega_k \gamma_0 + v (k_x \gamma_1 + k_y \gamma_2)\mright] \\
&\hspace{20pt} - \iu v_z (k_z \gamma_3) - \mu \gamma_0
\end{aligned} \label{eq:G-1-psi}
\end{align}
has a Dirac form.
Consequently, at high energies ($\abs{\omega_k} \gg \abs{\mu}$) the symmetry of the system is enhanced to $\SO(4)$ with generators $J_{\mu \nu} = - \tfrac{\iu}{4} [\gamma_{\mu}, \gamma_{\nu}]$.
The chemical potential $\mu$ breaks this symmetry down to $\SO(3)$: the group of spatial rotations which is generated by $L_i \defeq \tfrac{1}{2} \sum_{jk} \LCs_{ijk} J_{jk}$.
The neglected cubic term which is proportional to $\uptau_2 \Pauli_z = \iu \gamma_0 \gamma_5$, where $\gamma_5 \defeq \gamma_0 \gamma_1 \gamma_2 \gamma_3 = - \uptau_1 \Pauli_z$, reduces the symmetry group further down to the dihedral group generated by $C_{nz}$ and $C_{2 x}'$ that we started with.
Note how $P = \gamma_0$ and $\Theta = - \gamma_1 \gamma_3$, and how $L_3 = J_{12} = \tfrac{1}{2} \uptau_0 \Pauli_z$ and $L_1 = J_{23} = \tfrac{1}{2} \uptau_3 \Pauli_x$ agree with Eqs.~\eqref{eq:J12-exp} and~\eqref{eq:J31-exp}, respectively.

\begin{table}[t]
\caption{The symmetry classification of all $16$ matrices that can be constructed from the four $\gamma_{\mu}$ matrices.
Below $\gamma_5 \defeq \gamma_0 \gamma_1 \gamma_2 \gamma_3$, $L_i \defeq - \tfrac{\iu}{4} \sum_{jk} \LCs_{ijk} \gamma_j \gamma_k$, $i, j, k \in \{1, 2, 3\}$, $\ell$ is the angular momentum under $\SO(3)$ rotations which are generated by $L_i$, and $\hat{\mathcal{K}}$ is the complex conjugation operator.
Note that we're using a basis in which $\gamma_0^{*} = \gamma_0$, $\gamma_1^{*} = - \gamma_1$, $\gamma_2^{*} = \gamma_2$, and $\gamma_3^{*} = - \gamma_3$.
All the listed matrices are Hermitian.}
{\renewcommand{\arraystretch}{1.3}
\renewcommand{\tabcolsep}{8.5pt}
\begin{tabular}{c|ccc} \hline\hline
& $\hat{P} = \gamma_0$ & $\SO(3)$ rotations & $\hat{\Theta} = - \gamma_1 \gamma_3 \hat{\mathcal{K}}$ \\ \hline
$\one$, $\gamma_0$ & $+1$ & $\ell = 0$ & $+1$ \\
$\gamma_5$ & $-1$ & $\ell = 0$ & $+1$ \\
$\iu \gamma_0 \gamma_5$ & $-1$ & $\ell = 0$ & $-1$ \\
$\gamma_i$ & $-1$ & $\ell = 1$ & $+1$ \\
$\iu \gamma_0 \gamma_i$ & $-1$ & $\ell = 1$ & $-1$ \\
$L_i$, $\iu \gamma_i \gamma_5$ & $+1$ & $\ell = 1$ & $-1$
\\ \hline\hline
\end{tabular}}
\label{tab:gamma-class}
\end{table}

The alternative choice of Eq.~\eqref{eq:J31-exp-v2} as $C_{2 x}'$ would have given us the gamma matrices $\gamma_0' = \uptau_3 \Pauli_0$, $\gamma_1' = \uptau_1 \Pauli_x$, $\gamma_2' = \uptau_1 \Pauli_y$, and $\gamma_3' = \uptau_1 \Pauli_z$.
These are related to the previous ones through $\mathcal{B}^{\dag} \gamma_{\mu}' \mathcal{B} = \gamma_{\mu}$, where $\mathcal{B} = \diag(1,1,-\iu,\iu)$.
All later results rely only on the intrinsic Clifford algebra structure of the gamma matrices and their precise form in no way affects any of our conclusions.

We have thus found that anisotropic gapped Dirac models describe SOC-inverted bands of opposite parities near the $\Gamma$ point.
This is true for other high-symmetry points of the Brillouin zone as well if $P$, $C_{n z}$ with $n\geq 3$, and $C_{2 x}'$ are symmetry operations (belong to the little group) of these points.
When the high-symmetry points $\vb{k}_{*}$ have multiplicity higher than one, as happens when not all symmetry operations map $\vb{k}_{*}$ to $\vb{k}_{*}$ modulo $\vb{G}$, multiple valleys arise, each described by a Dirac model.
Although effective Dirac models have been found long ago in graphite, bismuth, and \ce{SnTe}~\cite{Wallace1947, Cohen1960, Wolff1964, Rogers1968, Hsieh2012, Ando2013, Ando2015}, and more recently in topological insulators such as \ce{Bi2Se3}, \ce{Bi2Te3}, and \ce{Sb2Te3}~\cite{Zhang2009, Liu2010}, the derivation of this section showcases that this generically holds true for band-inverted systems with SOC.

In light of the previously derived action~\eqref{eq:HS-int-action}, the part describing the internal dynamics of the plasmon field is given by
\begin{align}
\mathcal{S}_{\phi}[\phi] &= \frac{1}{2} \sum_q \phi(-q) V^{-1}(q) \phi(q),
\end{align}
where in the bare plasmon propagator
\begin{align}
V^{-1}(q) &= \epsilon_{\perp} \vb{q}_{\perp}^2 + \epsilon_z q_z^2 \label{eq:V-1-phi}
\end{align}
we allow for anisotropy.

Within the Dirac model, electric dipole moments are represented by $\psi^{\dag} \gamma_i \psi = \bar{\psi} \gamma_0 \gamma_i \psi$, where $i \in \{1, 2, 3\}$.
To see why, we note that the $\iu \gamma_0 \gamma_i$ which enter $H_{\vb{k}}$ transform as $\vb{k}$.
Therefore multiplying with $\iu \gamma_0$ will preserve the parity, while inverting the time-reversal sign, to give the unique Hermitian matrices which transform as electric dipoles; see Table~\ref{tab:gamma-class}.
Ferroelectric modes couple to Dirac fermions in the same way~\cite{KoziiBiRuhman2019, Kozii2022}, as expected from symmetry.
The electrostatic coupling term thus equals
\begin{align}
\mathcal{S}_{\psi\phi}[\psi, \phi] &= \frac{\iu}{\sqrt{\beta L^d}} \sum_{q} \phi(-q) \rho(q),
\end{align}
where $\beta = 1 / (k_B T)$, $L^d$ is the volume, and
\begin{align}
\rho(q) &= - \sum_k \bar{\psi}(k) A(k, k+q) \psi(k+q). \label{eq:rho-psi}
\end{align}
In the bare interaction vertex
\begin{align}
\begin{aligned}
A(k, p) &= e \gamma_0 + \iu \eta_{\perp} (k_x - p_x) \gamma_0 \gamma_1 \\
&\hspace{10pt} + \iu \eta_{\perp} (k_y - p_y) \gamma_0 \gamma_2 + \iu \eta_z (k_z - p_z) \gamma_0 \gamma_3
\end{aligned} \label{eq:A-psi}
\end{align}
we allow for anisotropy between the in-plane $\eta_{\perp}$ and out-of-plane $\eta_z$ electric dipole moments.
For later convenience, we retained the dependence of $A(k, p)$ on both the incoming $p$ and outgoing $k$ electron four-momenta.

\subsection{Polarization and optical conductivity} \label{sec:Dirac-opticond}
The polarization or plasmon self-energy $\Pi(q)$ is defined with the convention
\begin{align}
\mathscr{V}^{-1}(q) &= V^{-1}(q) + \Pi(q),
\end{align}
where $\mathscr{V}(q_1) \Kd_{q_1 + q_2} = \ev{\phi(q_1) \phi(q_2)}$ is the dressed plasmon propagator.
The small-momentum behavior of the polarization determines the symmetric part of the optical conductivity in the following way:
\begin{align}
\sigma_{ij}(\nu_q) &= - \iu \frac{\nu_q}{2} \mleft.\frac{\partial^2 \Pi^R(\nu_q, \vb{q})}{\partial q_i \partial q_j}\mright|_{\vb{q} = \vb{0}}.
\end{align}
Here, $\Pi^R(q) = \Pi^R(\nu_q, \vb{q})$ is the retarded real-time polarization which is obtained from $\Pi(q) = \Pi(\omega_q, \vb{q})$ via analytic continuation $\iu \omega_q \to \hbar \nu_q + \iu 0^+$.

Within RPA, $\Pi(q)$ is given by the fermionic polarization bubble which would have the form
\begin{align}
\Pi(q) \propto \sum_{\vb{k} n n' s s'} \frac{f_{\vb{k} + \vb{q} n} - f_{\vb{k} n'}}{\varepsilon_{\vb{k} + \vb{q} n} - \varepsilon_{\vb{k} n'} + \iu \omega_q} \abs{\braket{u_{\vb{k} + \vb{q} n s}}{u_{\vb{k} n' s'}}}^2
\end{align}
if we ignored the dipolar coupling.
Here, $\varepsilon_{\vb{k} n}$ are the dispersions, $u_{\vb{k} n s}$ the eigenvectors, and $f_{\vb{k} n} = 1 / \mleft(\Elr^{\beta \varepsilon_{\vb{k} n}} + 1\mright)$ are the Fermi-Dirac occupation factors.

In most systems, the electric monopole-monopole contribution to $\Pi(q)$, which is schematically written above, is dominant and gives the leading contribution to the optical conductivity.
However, in quasi-2D systems the Hamiltonian $H_{\vb{k}}$ has weak $k_z$-dependence, making both $\varepsilon_{\vb{k} n}$ and $u_{\vb{k} n s}$ weakly dependent on $k_z$, in contrast to the coupling of the $z$-axis electric dipoles $\eta_z$ [Eq.~\eqref{eq:A-psi}].
It then follows that the monopole-monopole contribution to $\sigma_{zz}(\nu_q)$ is small in quasi-2D systems, whereas the dipolar contributions can be large.
In particular, for the model of the previous section we have evaluated the polarization in the quasi-2D limit:
\begin{align}
v_z &= 0, & \eta_{\perp} &= 0,
\end{align}
which is also of interest for RG reasons discussed in the next section.
The $T = 0$ result is (Appendix~\ref{sec:polarization}):
\begin{align}
\begin{aligned}
\Pi^R(\nu_q, 0, q_z) &= \frac{\Lambda_z m^2 \eta_z^2 q_z^2}{\pi^2 v^2 \hbar \nu_q} \bigg[\log\abs{\frac{2 \mu + \hbar \nu_q}{2 \mu - \hbar \nu_q}} \\
&\hspace{70pt} + \iu \pi \Theta\mleft(\hbar \abs{\nu_q} - 2 \mu\mright)\bigg],
\end{aligned} \label{eq:Pi-R-quasi2D}
\end{align}
where $\Lambda_z$ is the $q_z$ cutoff, $q_z \in [- \Lambda_z, \Lambda_z]$, $\mu = \sqrt{m^2 + v^2 k_F^2}$ is the chemical potential, $\Theta$ is the Heaviside theta function.
Note that in the no doping limit $k_F \to 0$, $\mu$ should go to $m$, not $0$, in the above expression.
The $z$-axis optical conductivity is thus exclusively given by the $z$-axis dipole fluctuations:
\begin{align}
\begin{aligned}
\sigma_{zz}(\nu_q) &= \frac{\Lambda_z m^2 \eta_z^2}{\pi^2 v^2 \hbar} \bigg(\pi \Theta\mleft(\hbar \abs{\nu_q} - 2 \mu\mright) \\[-2pt]
&\hspace{70pt} - \iu \log\abs{\frac{2 \mu + \hbar \nu_q}{2 \mu - \hbar \nu_q}}\bigg).
\end{aligned}
\end{align}
Due to interband excitations, above the gap we obtain a flat real part of the conductivity, which is very similar to the usual behavior of the in-plane optical conductivity for a two-dimensional Dirac spectrum~\cite{Gorbar2002, Pyatkovskiy2008}.
The surprise is that we obtain this result for the $z$-axis conductivity, even though the band velocity along this direction is zero.
The matrix element responsible for this is exclusively the anomalous dipole element of Eq.~\eqref{eq:dipol-matrix-elem}.

In conclusion, in quasi-2D Dirac systems the $z$-axis dipole fluctuations that are so important for our pairing mechanism of Sec.~\ref{sec:Dirac-pairing} are directly observable in the $z$-axis optical conductivity.

\subsection{Renormalization group (RG) analysis} \label{sec:Dirac-RG}
\begin{figure}[t]
\includegraphics[width=0.85\columnwidth]{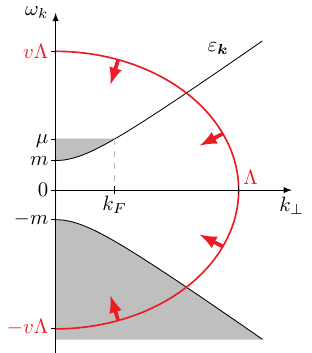}
\caption{A schematic of the RG procedure.
Here, $\omega_k$ and $k_{\perp} = \sqrt{k_x^2 + k_y^2}$ are the frequency and in-plane momentum, $2 m$ is the band gap, $\mu$ is the chemical potential, $k_F$ is the Fermi wavevector, and $\varepsilon_{\vb{k}} = \sqrt{m^2 + v^2 \vb{k}_{\perp}^2}$ is the dispersion.
The occupied states are shaded in grey and the cutoff of Eq.~\eqref{eq:cutoff-def} is highlighted in red.
Arrows indicate the direction of the RG flow.}
\label{fig:RG-sketch}
\end{figure}

Here we study how the fluctuations of high-energy states modify the low-energy physics of our model.
To this end, we first analyze the naive scaling under RG flow, which is depicted in Fig.~\ref{fig:RG-sketch}.
We show that the electric dipole coupling is irrelevant in 3D systems, while in quasi-2D systems its out-of-plane component is marginal.
Afterwards, for the quasi-2D case we derive the $1$-loop RG flow equations in the limit of a large number of fermionic flavors $N$ and we establish that the out-of-plane coupling $\eta_z$ is marginally relevant.
Consequently, $\eta_z$ becomes enhanced at low energies.

Cooper pairing, which we study in the next section, takes place only when the screening is strong enough.
The Thomas-Fermi wavevector $k_{\text{TF}} = \sqrt{e^2 g_F / \epsilon_0}$ thus needs to be larger than the Fermi sea size $k_F$.
Since the density of states $g_F \propto k_F^2 / (\hbar v_F)$, $k_{\text{TF}} \propto k_F \sqrt{\alpha}$ where $\alpha = e^2 / (\hbar v_F \epsilon_0)$ is the monopole coupling constant.
For this reason, throughout this section we focus on the strong-coupling regime $\alpha \gg 1$.

The strong-coupling regime is not accessible through direct perturbation theory, which is why we use a large-$N$ expansion, $N$ being the number of fermion flavors.
Formally, we modify the model by introducing an additional summation over flavor indices in Eqs.~\eqref{eq:S-psi} and~\eqref{eq:rho-psi}.
Although in the end we take $N$ to be of order unity, the hope is that by organizing the calculation in orders of $1/N$ we can at least make definite statements about some strongly coupled model that resembles our model.
When the band inversion point is not located at $\vb{k} = 0$, multiple valleys arise, each described by a Dirac model.
This naturally gives larger values for $N$, provided that the inter-valley interactions are small.

At the start of the RG procedure, the momentum cutoff $\Lambda$ is initially much larger than the Fermi wave vector $k_F$ and we integrate out high-energy degrees of freedom until $\Lambda$ becomes comparable to $k_F$; see Fig.~\ref{fig:RG-sketch}.
To a first approximation, we may thus set the chemical potential mid-gap, i.e., $k_F$ to zero.
Since we are only interested in the low-temperature physics, we may also set $T = 0$.
Throughout this section, we thus set
\begin{align}
\mu &= 0, & T &= 0. \label{eq:muT0-limit}
\end{align}
Finite $\mu$ and $T$ are both reintroduced later when we study Cooper pairing given a cutoff $\Lambda \sim k_F$.

First, we study the tree-level scaling (when $\mu = k_F = T = 0$).
In light of the Dirac form, the cutoff $\Lambda$ we impose on both momenta and frequencies according to
\begin{align}
\abs{k}^2 \defeq \omega_k^2 / v^2 + k_x^2 + k_y^2 + (v_z / v)^2 k_z^2 < \Lambda^2. \label{eq:cutoff-def}
\end{align}
We decompose the fields $\psi = \psi_< + \psi_>$ and $\phi = \phi_< + \phi_>$  into slow and fast parts with four-momenta within $0 < \abs{k} < \Lambda/b$ and $\Lambda/b < \abs{k} < \Lambda$, respectively; here $b = \Elr^{\ell} > 1$.
The naive slow-field action, which is obtained by substituting the slow fields into Eq.~\eqref{eq:model-action}, can be rescaled into the original action via
\begin{align}
\begin{aligned}
k &\mapsto k' = b^1 k, \\
\psi(k) &\mapsto \psi'(k') = b^{-\eta_{\psi}} \psi(k), \\
\phi(q) &\mapsto \phi'(q') = b^{-\eta_{\phi}} \phi(q),
\end{aligned}
\end{align}
where we choose $\eta_{\psi}$ and $\eta_{\phi}$ so that the fermionic frequency $\bar{\psi}(k) \omega_k \psi(k)$ and monopole coupling $\phi(-q) \psi(k) e \gamma_0 \psi(k+q)$ terms are invariant.
The naive scaling of the various model parameters is in 3+1D given by ($\eta_{\psi} = 5/2$, $\eta_{\phi} = 3$)
\begin{align}
\begin{aligned}
m' &= b^1 m, &&& v' &= v, &&& e' &= e,  \\
\epsilon_{\perp}' &= \epsilon_{\perp}, &&& \epsilon_z' &= \epsilon_z, \\
\eta_{\perp}' &= b^{-1} \eta_{\perp}, &&& \eta_z' &= b^{-1} \eta_z.
\end{aligned}
\end{align}
The electric dipole coupling is naively irrelevant, as are all higher-order momentum-conserving local terms in the action which preserve $\phi \to - \phi$ symmetry and particle number.
Because the scaling of $\eta_{\perp}$ and $\eta_z$ only receives loop corrections of order $N^{-1}$ or higher, in 3D Dirac systems electric dipole moments become increasingly weak at low energies.

In quasi-2D systems, however, $v_z \approx 0$ and the Fermi surface is cylindrical instead of spherical.
Consequently, during the RG we do not rescale the momenta along $z$.
This changes the naive scaling dimensions to ($\eta_{\psi} = 2$, $\eta_{\phi} = 2$)
\begin{align}
\begin{aligned}
\epsilon_{\perp}' &= b^{-1} \epsilon_{\perp}, &&&\qquad
\epsilon_z' &= b^{1} \epsilon_z, \\
\eta_{\perp}' &= b^{-1} \eta_{\perp}, &&&\qquad
\eta_z' &= \eta_z.
\end{aligned}
\end{align}
Hence the out-of-plane dipole moment is now  marginal, and we shall later see that loop corrections make it marginally relevant.
The monopole coupling $e$ remains marginal.
Intuitively, the reason why the dipolar couplings $\eta_{\perp}$ and $\eta_z$ were previously irrelevant is that they come with an additional power of momentum compared to the charge $e$.
As this momentum becomes smaller because of the restricted momentum range ($\Lambda \to \Lambda/b$), they become increasingly less important, at least in three dimensions for $k_F \ll \Lambda$.
In quasi-2D systems, however, the exchanged momentum along the $\vu{z}$ direction is always large (on the order of the Brillouin zone height) and the importance of the $\eta_z$ term is always (naively) the same, which explains its marginality as $\Lambda$ is decreased.
As for $\epsilon_z$, it is relevant, as expected for what is essentially a $z$-dependent mass of the plasmon field.
However, the electrons themselves also gap the plasmon field and in the strong screening limit their contribution is dominant.
This is why we do not consider the flow of $\epsilon_z$ later on.

Given our interest in dipole effects, we focus on quasi-2D systems.
Since $\eta_{\perp}$ is irrelevant, we may set it to zero from the outset.
We therefore consider the regime
\begin{align}
v_z &= 0, & \eta_{\perp} &= 0 \label{eq:quasi-2D-param}
\end{align}
from now on.
In practice, the $z$-axis dispersion and $\eta_{\perp}$ have to be small compared to $m$ and $\eta_z$, respectively, for our calculation to apply.
For quasi-2D geometries, we shall find it convenient to use bolded vectors with $\perp$ subscripts to denote in-plane vectors.
For instance:
\begin{align}
\begin{aligned}
k &= (\omega_k, \vb{k}_{\perp}, k_z), &&&\qquad
\vb{k}_{\perp} &= (k_x, k_y), \\
q &= (\omega_q, \vb{q}_{\perp}, q_z), &&&\qquad
\vb{q}_{\perp} &= (q_x, q_y).
\end{aligned} \label{eq:bold-vec-conv}
\end{align}

To formulate the RG flow equations, we use the Callan-Symanzik equations~\cite{Zinn-Justin2002}.
Let us assume that we have found how all the states up to the cutoff $\Lambda$ renormalize the fermionic Green's function $G(k)$ of Eq.~\eqref{eq:G-1-psi} into $\mathscr{G}(k)$:
\begin{align}
\begin{aligned}
\mathscr{G}^{-1}(k) &= Z_m m \one - \iu Z_{\omega} \omega_k \gamma_0 \\
&\hspace{20pt} - \iu Z_v v (k_x \gamma_1 + k_y \gamma_2) + \cdots \, ,
\end{aligned}
\end{align}
and the same for the interaction vertex $A(k, p) \to \mathscr{A}(k, p)$ of Eq.~\eqref{eq:A-psi}:
\begin{align}
\mathscr{A}(k, p) = Z_e e \gamma_0 + \iu Z_{\eta z} \eta_z (k_z - p_z) \gamma_0 \gamma_3 + \cdots \, .
\end{align}
The Callan-Symanzik equations follow from the requirement that this asymptotic behavior for small $k, p$ stays preserved as we change $\Lambda$.
Before imposing this, we need to fix the scale of the fields $\psi$ and $\phi$ which can in general depend on $\Lambda$.
We choose $\psi \to Z_{\omega}^{-1/2} \psi$, in which case the Callan-Symanzik equations take the form:
\begin{align}
\begin{aligned}
\dv{}{\Lambda} \mleft[\frac{Z_m}{Z_{\omega}} m\mright] &= 0, &&&\quad
\dv{}{\Lambda} \mleft[\frac{Z_v}{Z_{\omega}} v\mright] &= 0, \\
\dv{}{\Lambda} \mleft[\frac{Z_e}{Z_{\omega}} e\mright] &= 0, &&&\quad
\dv{}{\Lambda} \mleft[\frac{Z_{\eta z}}{Z_{\omega}} \eta_z\mright] &= 0.
\end{aligned}
\end{align}
Because $\phi$ couples to the Noether charge of the $\Ugp(1)$ phase rotation symmetry $\psi \to \Elr^{\iu \vartheta} \psi$, there is an exact Ward identity $Z_e = Z_{\omega}$ which implies that the charge $e$ does not flow; see Appendix~\ref{sec:Ward-id}.
As for the other parameters, the chain rule gives the RG flow equations:
\begin{align}
\sum_j \mleft(\Kd_{ij} + \frac{g_j}{Z_i} \pdvc{Z_i}{g_j}{\Lambda, g_{\ell}}\mright) \frac{\Lambda}{g_j} \dv{g_j}{\Lambda} &= - \frac{\Lambda}{Z_i} \pdvc{Z_i}{\Lambda}{g_{\ell}},
\end{align}
where
\begin{align}
g_i &= \begin{pmatrix}
m \\ v \\ \eta_z
\end{pmatrix}, &
Z_i &= \begin{pmatrix}
{Z_m}/{Z_{\omega}} \\
{Z_v}/{Z_{\omega}} \\
{Z_{\eta z}}/{Z_{\omega}}
\end{pmatrix}.
\end{align}
Since $Z_i = 1 + \bigO(N^{-1})$, as we later show, to $N^{-1}$ order the RG flow equations simplify to:
\begin{align}
\frac{\Lambda}{g_i} \dv{g_i}{\Lambda} &= - \frac{\Lambda}{Z_i} \pdv{Z_i}{\Lambda}. \label{eq:RG-flow-expr}
\end{align}
In these RG flow equations we have not included $\epsilon_{\perp}$ or $\epsilon_z$ because the bare interaction is negligible compared to the polarization in the strong coupling limit.
$\epsilon_{\perp}$ and $\epsilon_z$ we shall therefore keep constant (independent of $\Lambda$) and only include in various expressions to make them dimensionless.

To lowest order in $N$, the plasmon self-energy $\Pi(q)$ is given by the fermionic polarization bubble which equals (Appendix~\ref{sec:polarization}):
\begin{align}
\Pi(q) &= N \Lambda_z \frac{\vb{q}_{\perp}^2 (e^2 + \eta_z^2 q_z^2)}{4 \pi^2 \sqrt{\omega_q^2 + v^2 \vb{q}_{\perp}^2}} \mleft[(1-r_q^2) \arcctg r_q + r_q\mright] \notag \\
&\hspace{20pt} + N \Lambda_z \frac{2 m^2 \eta_z^2 q_z^2}{\pi^2 v^2 \sqrt{\omega_q^2 + v^2 \vb{q}_{\perp}^2}} \arcctg r_q, \label{eq:mu0-polarization}
\end{align}
where $\Lambda_z$ is the $q_z$ cutoff, $q_z \in [- \Lambda_z, \Lambda_z]$, and
\begin{align}
r_q \defeq \frac{2 m}{\sqrt{\omega_q^2 + v^2 \vb{q}_{\perp}^2}}.
\end{align}
Notice how $\Pi(q)$, unlike the bare $V^{-1}(q)$ of Eq.~\eqref{eq:V-1-phi}, is frequency-dependent and non-analytic at $q = 0$.

The next step is to evaluate the various renormalization factors $Z_i$, which we do to $N^{-1}$ order.
The relevant self-energy and vertex correction diagrams are standard and the details of their evaluation are delegated to Appendix~\ref{sec:RG-diagrams}.
Although the shell integrals cannot be carried out analytically, they can be simplified by introducing the dimensionless parameters:
\begin{align}
\begin{aligned}
\tilde{m} &\defeq \frac{m}{v \Lambda}, &&&\qquad
\alpha &\defeq \frac{e^2}{\epsilon_{\perp} v}, \\
\tilde{\eta}_z &\defeq \frac{\Lambda_z \eta_z}{e}, &&&\qquad
\lambda &\defeq \frac{\Lambda}{\Lambda_z},
\end{aligned}
\end{align}
and expressing the shell momentum $q = (\omega_q, \vb{q}_{\perp}, q_z)$ in terms of dimensionless $x, z$ through $\omega_q = v \Lambda x$, $\abs{\vb{q}_{\perp}} = \Lambda \sqrt{1-x^2}$, and $q_z = \Lambda_z z$.

\begin{widetext}
The strong-coupling large-$N$ RG flow equations are to $1$-loop order:
\begin{align}
\begin{aligned}
\frac{1}{\tilde{m}} \dv{\tilde{m}}{\ell} &= \beta_{m} = 1 + \frac{4}{\pi^2 N (1+\tilde{m}^2)^2} \int_0^1 \dd{z} \int_0^1 \dd{x} \frac{B_{m}(x, z)}{P(x, z)}, \\
\frac{1}{\alpha} \dv{\alpha}{\ell} &= \beta_{\alpha} = - \frac{4}{\pi^2 N (1+\tilde{m}^2)^2} \int_0^1 \dd{z} \int_0^1 \dd{x} \frac{B_{\alpha}(x, z)}{P(x, z)}, \\
\frac{1}{\tilde{\eta}_z} \dv{\tilde{\eta}_z}{\ell} &= \beta_{\eta z} = \frac{4}{\pi^2 N (1+\tilde{m}^2)^2} \int_0^1 \dd{z} \int_0^1 \dd{x} \frac{B_{\eta z}(x, z)}{P(x, z)}, \\
\frac{1}{\lambda} \dv{\lambda}{\ell} &= -1,
\end{aligned} \label{eq:RG-flow-final}
\end{align}
where $\ell$ determines the cutoff through $\Lambda = \Lambda_0 / b = \Lambda_0 \Elr^{- \ell}$ and
\begin{align}
\begin{aligned}
B_{m}(x, z) &= (1-x^2) \mleft[1 - (1+\tilde{m}^2) \tilde{\eta}_z^2\mright] \lambda^2 - (x^2 + \tilde{m}^2) z^2 \tilde{\eta}_z^2, \\
B_{\alpha}(x, z) &= (1-x^2 + 2\tilde{m}^2) (\lambda^2 + z^2 \tilde{\eta}_z^2), \\
B_{\eta z}(x, z) &= 2 \tilde{m}^2 (\lambda^2 + z^2 \tilde{\eta}_z^2), \\
P(x, z) &= (1-x^2) \tfrac{2}{\pi} \mleft[(1-4\tilde{m}^2) \arcctg(2\tilde{m}) + 2\tilde{m}\mright] \lambda^2 \\
&\hspace{20pt} + \mleft\{(1-x^2) \tfrac{2}{\pi} \mleft[(1+4\tilde{m}^2) \arcctg(2\tilde{m}) + 2\tilde{m}\mright] + 8 x^2 \tilde{m}^2 \tfrac{2}{\pi} \arcctg(2\tilde{m})\mright\} z^2 \tilde{\eta}_z^2.
\end{aligned}
\end{align}
These RG flow equations are the main result of this section.
\end{widetext}

By inspection, one sees that $P$, $B_{\alpha}$, and $B_{\eta z}$ are strictly positive for all $x$ and $z$, whereas $B_{m}$ can be positive or negative.
Consequently, the dimensionless out-of-plane electric dipole moment $\tilde{\eta}_z$ is always marginally relevant, while the effective fine-structure constant $\alpha$ is always marginally irrelevant.

\begin{figure}[b!]
\includegraphics[width=\columnwidth]{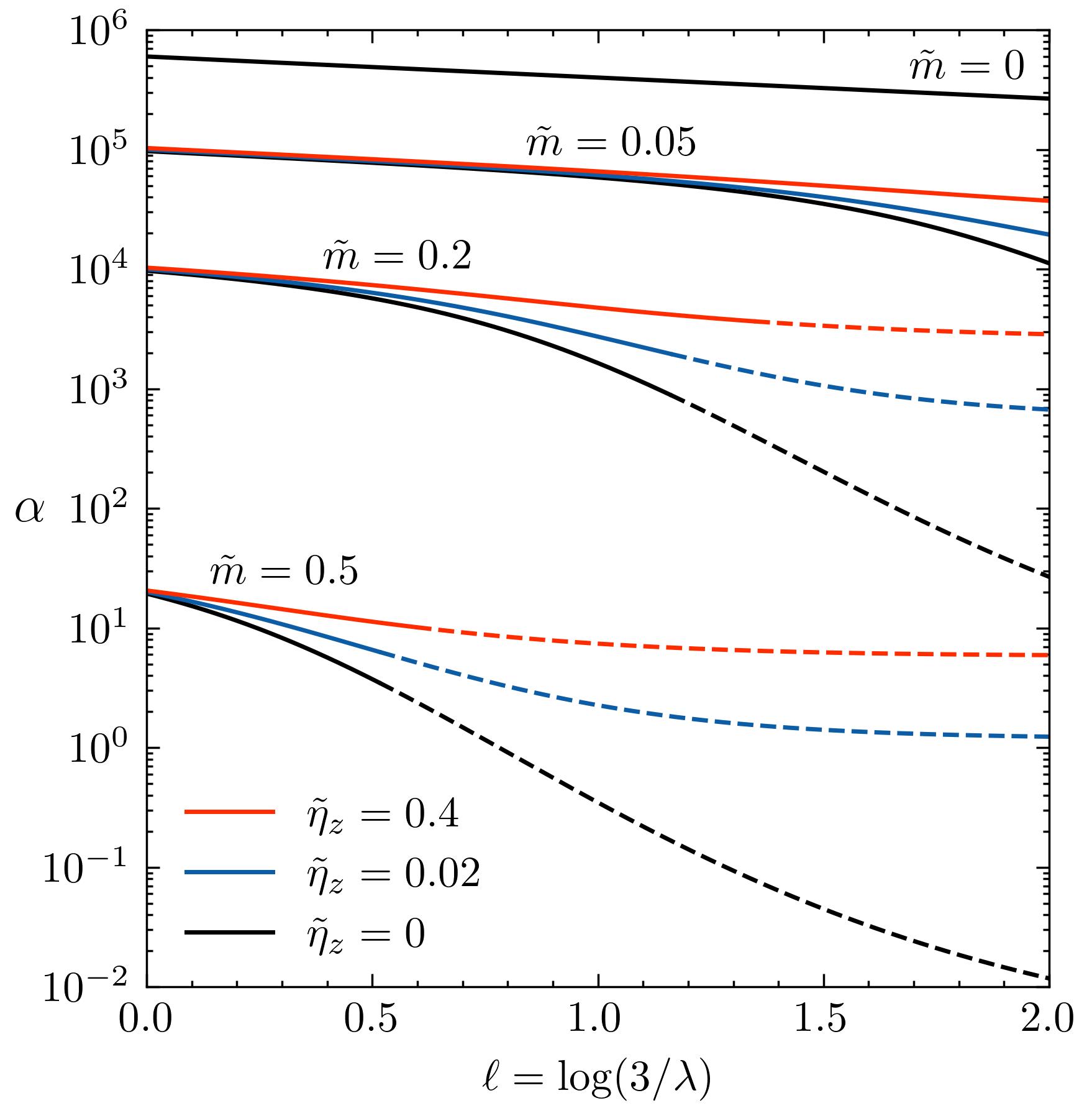}
\caption{The RG flow of $\alpha$ with $N = 1$ for various initial $\tilde{m}(\ell=0)$ and $\tilde{\eta}_z(\ell=0)$, as indicated on the figure.
Solid lines become dashed when $\tilde{m}(\ell) > 1$.
The $\alpha(\ell)$ curves associated with different initial masses we have offset relative to each other via multiplication (displacement on a log scale).
We are allowed to do this because $\alpha(\ell=0)$ enters as a multiplicative factor in the solution of the RG flow equations~\eqref{eq:RG-flow-final}.}
\label{fig:RG-alpha-flow}
\end{figure}

The flow of the dimensionless gap $\tilde{m}$ is the simplest: it grows with an exponent that approximately equals $+1$ even when we extrapolate $N \to 1$, as the numerical evaluating of the shell integral shows.
Once $\tilde{m}$ becomes on the order of $\sim 1$, the RG flow should be terminated.
Even though large $\tilde{m}$ are thus never reached, let us nonetheless note that all three $B_i / P \propto \tilde{m}$ for large $\tilde{m}$ and therefore the flow of both $\alpha$ and $\tilde{\eta}$ is suppressed as $\tilde{m} \to + \infty$, as expected.
In addition, the RG flow equations are symmetric with respect to $\tilde{m} \to - \tilde{m}$ so we may always choose $\tilde{m} \geq 0$, as we do below.

The flow of $\alpha$ for a gapless 2D Dirac system without electric dipoles was analyzed in Ref.~\cite{DTSon2007} and we recover their $\partial_{\ell} \alpha = - \frac{4}{\pi^2 N} \alpha$ result when we set $\tilde{m} = 0$.
Our analysis shows that the flow towards small $\alpha$ persists for finite gaps $\tilde{m}$ and finite $z$-axis dipolar couplings $\tilde{\eta}_z$.
The detailed behavior is shown in Fig.~\ref{fig:RG-alpha-flow}, where we plot the flow of $\alpha$ for different initial values of the mass $\tilde{m}$ and dipole element $\tilde{\eta}_z$.
Notice that $\alpha$ does not enter any of the beta functions $\beta_i$ in the strong-coupling limit $\alpha \to \infty$.
Hence, we may offset the solutions via multiplication, as we did in Fig.~\ref{fig:RG-alpha-flow} for illustration purposes only.
The suppression of $\alpha$ is stronger for intermediate $\tilde{m} \sim 1$ than for very small $\tilde{m} \to 0$, and we shall later see that this is accompanied by an enhancement of $\tilde{\eta}$ that also predominantly takes place for $\tilde{m} \sim 1$.
On the other hand, because $B_{\alpha} / P = 1$ when $\tilde{m} = 0$, $\tilde{\eta}_z$ has a negligible effect on the flow of $\alpha$ for small $\tilde{m}$.
For intermediate $\tilde{m} \sim 1$, small $\tilde{\eta}_z$ are more favorable for the suppression of $\alpha$ than large $\tilde{\eta}_z$, as can be seen from Fig.~\ref{fig:RG-alpha-flow}.
Both positive and negative $\tilde{\eta}_z$ affect $\alpha$ the same way because of horizontal reflection symmetry $\tilde{\eta}_z \to - \tilde{\eta}_z$, which is respected by Eqs.~\eqref{eq:RG-flow-final}; below we assume $\tilde{\eta}_z \geq 0$.

\begin{figure}[t]
\includegraphics[width=\columnwidth]{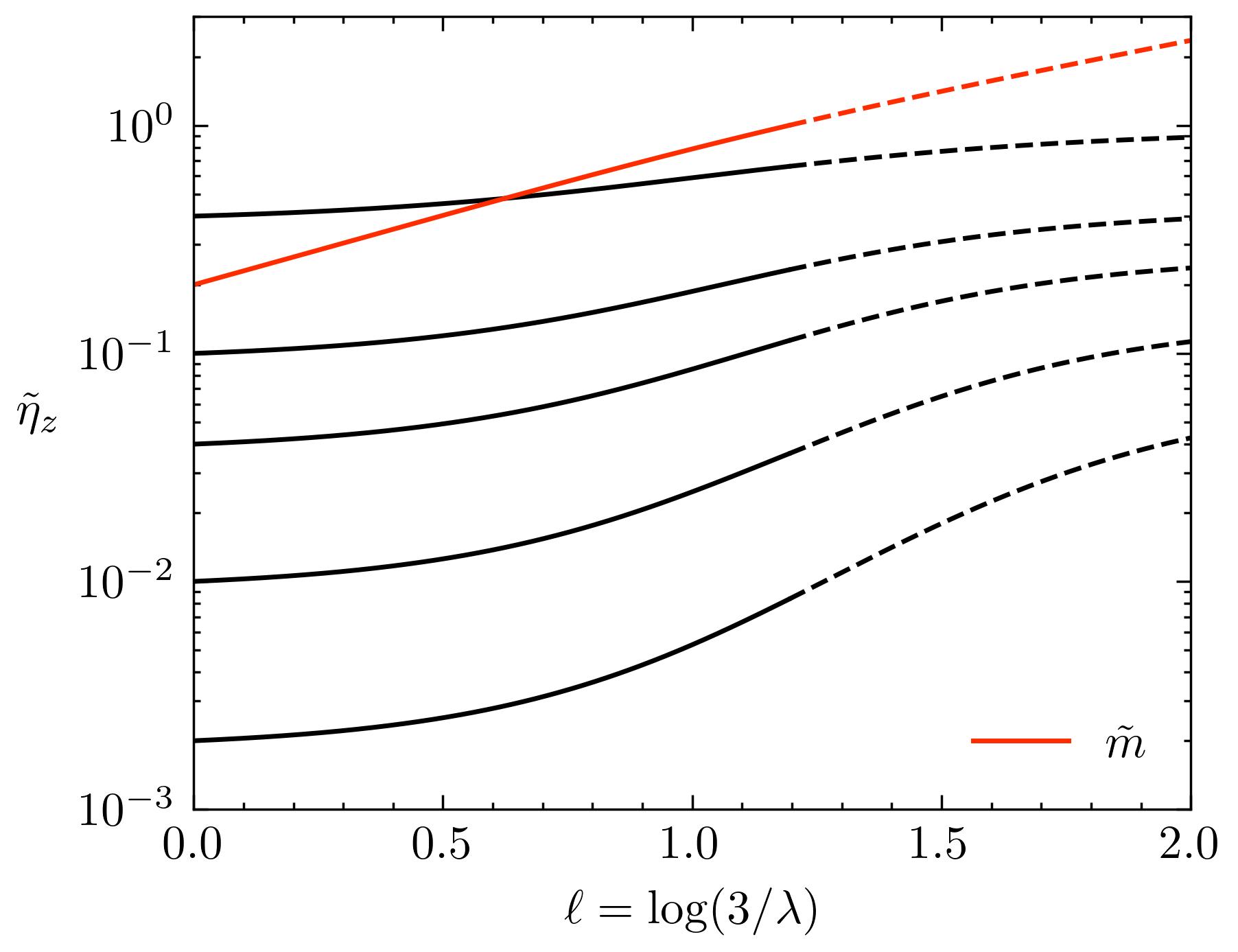}
\caption{The RG flow of $\tilde{\eta}_z$ and $\tilde{m}$ with $N = 1$ for an initial $\lambda(\ell = 0) = 3$, $\tilde{m}(\ell = 0) = 0.2$, and $\tilde{\eta}_z(\ell = 0) \in \{0.002, 0.01, 0.04, 0.1, 0.4\}$.
Solid lines become dashed when $\tilde{m}(\ell) > 1$.
There are small variations in how $\tilde{m}$ flows, depending on $\tilde{\eta}_z(\ell = 0)$, which we are not shown.
The same scale is used for both $\tilde{m}$ and $\tilde{\eta}_z$.}
\label{fig:RG-etaz-flow}
\end{figure}

The dependence of the flow of the dipole strength $\tilde{\eta}_z$ on the mass $\tilde{m}$ is more subtle than that of the monopole coupling $\alpha$.
Its beta function $\beta_{\eta z}$ vanishes for both small and large $\tilde{m}$.
That large gaps suppress the flow of $\tilde{\eta}_z$ is expected because large gaps suppress the mixing of parities that is needed for high-energy fluctuations to affect electric dipole moments.
Less obvious is the fact that there is a chiral $\Ugp(1)$ symmetry $\psi \to \Elr^{\iu \vartheta \gamma_3} \psi$ in the gapless limit $m \to 0$ (with $k_F = v_z = \eta_{\perp} = 0$) and that the out-of-plane electric dipole moments precisely couple to its charge $\bar{\psi} \gamma_0 \gamma_3 \psi$.
As a result, the associated Ward identity guarantees that $Z_{\eta z} = Z_{\omega}$, precluding any renormalization of $\eta_z$, as we prove in Appendix~\ref{sec:Ward-id}.
The largest increase in $\tilde{\eta}_z$ thus happens for moderate $\tilde{m} \sim 1$, and for large $\lambda$, as follows from the fact that $B_{\eta z} \propto \lambda^2$.

The numerical results for the flow of the $z$-axis dipole element $\tilde{\eta}_z$ are shown in Fig.~\ref{fig:RG-etaz-flow}.
These results depend on the initial values of $\lambda$, $\tilde{m}$, and $\tilde{\eta}_z$, which are specified below.
Note that they do not depend on $\alpha$ as long as it is large because $\alpha(\ell)$ decouples from the rest in the strong-coupling limit described by Eqs.~\eqref{eq:RG-flow-final}.

For $\lambda$, we assume that initially $\lambda = 3$, which corresponds to a reasonable amount of anisotropy for a quasi-2D system ($\Lambda = 3 \Lambda_z$).
The RG flow we run until $\ell = 2$, at which point $\lambda = 3 \Elr^{-2} = 0.41$.
The Fermi radius $k_F$, which we neglected [Eq.~\eqref{eq:muT0-limit}], is thus on the order of $0.2 \Lambda_z$.

Regarding the gap, in Fig.~\ref{fig:RG-alpha-flow} we only show the results for an initial $\tilde{m} = 0.2$.
We have explored other initial values as well and we have found that the enhancement of $\tilde{\eta}_z$ is comparable in magnitude to that shown in Fig.~\ref{fig:RG-alpha-flow} in the range $\tilde{m} \in \langle 0.05, 1.0\rangle$, whereas outside of this range it is a lot smaller.
As already remarked, the flow of $\tilde{m}$, given an initial value, is not significantly affected by $\tilde{\eta}_z$ so only one curve for $\tilde{m}(\ell)$ is shown in Fig.~\ref{fig:RG-alpha-flow}.

The RG flow is given for five different initial values of $\tilde{\eta}_z$, ranging from $0.002$ to $0.4$.
As can be seen in Fig.~\ref{fig:RG-alpha-flow}, although smaller $\tilde{\eta}_z$ tend to get more enhanced, sometimes by even two orders of magnitude (if we take $N \to 1$), the final value of $\tilde{\eta}_z(\ell=2)$ declines with decreasing $\tilde{\eta}_z(\ell=0)$.
Larger microscopic electric dipole moments $\tilde{\eta}_z(\ell=0)$ thus always lead to larger effective dipole moments $\tilde{\eta}_z(\ell=2)$.
It is also worth noting that the increase in $\tilde{\eta}_z$ is finite even if we extend $\ell$ to go from $-\infty$ to $+\infty$.
The reason lies in the fact discussed earlier that both small and large $\tilde{m}$ suppress the beta function of $\tilde{\eta}_z$.
Hence the dipole matrix element grows only in an intermediate window before $\tilde{m}$ becomes too large.
This should be contrasted to the flow of $\alpha$ which stops for large $\ell$, but is exponential for small $\ell \to - \infty$.

\section{Pairing due to electric monopole-dipole interactions} \label{sec:pairing}
The strongly repulsive nature of the Coulomb interaction is often one of the biggest obstacles to the formation of Cooper pairs.
Its monopole-monopole part by itself is repulsive and suppresses pairing.
However, the monopole-dipole and dipole-dipole parts can yield unconventional superconductivity if the screening and dipole moments are strong enough, as we show here.
Starting from an effective instantaneous interaction among Fermi-level electrons, such as the one obtained at the end of the RG flow of the previous section, we first summarize the formalism for analyzing superconducting instabilities.
Using this formalism, we then study the pairing due to electric monopole-dipole interactions for general systems and we derive a number of its properties.
Although we call this pairing after the monopole-dipole term only, we are not neglecting dipole-dipole interactions in our analysis, but are rather emphasizing the fact that the monopole-dipole coupling is the main source of pairing.
The pairing in quasi-2D Dirac metals, which were the subject of Sec.~\ref{sec:Dirac}, we analyze in the next section.

\subsection{Gap equation and formalism} \label{sec:lin-gap-eq}
To study Cooper pairing, we use the linearized gap equation.
If we keep the electron-electron interaction generic for the moment, we have
\begin{align}
\begin{aligned}
H_{\text{int}} = \frac{1}{4 L^d} &\sum_{1234} U^{\alpha_{1} \alpha_{2}}_{\alpha_{3} \alpha_{4}}(\vb{k}_{1},\vb{k}_{2},\vb{k}_{3},\vb{k}_{4}) \\[-3pt]
&\hspace{14pt} \times \psi_{\alpha_1}^{\dag}(\vb{k}_1) \psi_{\alpha_2}^{\dag}(\vb{k}_2) \psi_{\alpha_4}(\vb{k}_4) \psi_{\alpha_3}(\vb{k}_3),
\end{aligned}
\end{align}
where $U$ is fully antisymmetrized with respect to particle exchange.
At leading order in this interaction, one obtains the following linearized gap equation, formulated as an eigenvalue problem,
\begin{align}
\sum_n \int\limits_{\varepsilon_{\vb{k} n} = 0} \frac{\dd{S_{\vb{k}}}}{(2\pi)^d} \sum_{a=0}^{3} W_{ba}(\vb{p}_m, \vb{k}_n) d_{a}(\vb{k}_n) = \lambda \, d_{b}(\vb{p}_m). \label{eq:lin-gap-eq}
\end{align}
Here $n, m$ are band indices, $\varepsilon_{\vb{k} n}$ is the band dispersion displaced by the chemical potential, the momenta $\vb{k}_n, \vb{p}_m$ are on the Fermi surfaces which are determined by $\varepsilon_{\vb{k} n} = \varepsilon_{\vb{p} m} = 0$, $\dd{S_{\vb{k}}}$ is a surface element, $a = b = 0$ corresponds to even- and $a, b \in \{1, 2, 3\}$ to odd-parity pairing, $d_{a}(\vb{k}_n)$ is the pairing $\vb{d}$-vector, and $W_{ba}(\vb{p}_m, \vb{k}_n)$ is the pairing interaction.
This linearized gap equation applies to spin-orbit-coupled Fermi liquids with space- and time-reversal symmetries whose Fermi surfaces do not touch each other or have Van Hove singularities on them.

Positive pairing eigenvalues $\lambda$ correspond to superconducting states with transition temperatures:
\begin{align}
k_ B T_c &= \frac{2 \Elr^{\gamma_E}}{\pi} \hbar \omega_c \, \Elr^{- 1 / \lambda} \approx 1.134 \, \hbar \omega_c \, \Elr^{- 1 / \lambda}, \label{eq:BCS-Tc-expr}
\end{align}
where $\gamma_E$ is the Euler-Mascheroni constant and $\hbar \omega_c$ is the energy cutoff, which is much smaller than the bandwidth.
The leading instability is determined by the largest positive $\lambda$.

The pairing interaction is given by:
\begin{align}
W_{ba}(\vb{p}_m, \vb{k}_n) &= - \frac{1}{4 \abs{\grad_{\vb{p}} \varepsilon_{\vb{p} m}}^{1/2} \abs{\grad_{\vb{k}} \varepsilon_{\vb{k} n}}^{1/2}} \\
&\hspace{20pt} \times \hspace{-8pt} \sum_{\alpha_1 \alpha_2 \alpha_3 \alpha_4} \left[\Theta^{*} P_{\vb{p} m}^{b}\right]_{\alpha_2 \alpha_1} \left[P_{\vb{k} n}^{a} \Theta^{\intercal}\right]_{\alpha_3 \alpha_4} \notag \\[-3pt]
&\hspace{70pt} \times U^{\alpha_1 \alpha_2}_{\alpha_3 \alpha_4}(\vb{p}, -\vb{p}, \vb{k}, -\vb{k}), \notag
\end{align}
where $P_{\vb{k} n}^{a}$ are the band projectors:
\begin{align}
P_{\vb{k} n}^{a} = \sum_{s s'} u_{\vb{k} n s} (\Pauli_a)_{ss'} u_{\vb{k} n s'}^{\dag}.
\end{align}
Here $s, s' \in \{\uparrow, \downarrow\}$ are the pseudospins, $\Pauli_a$ are the Pauli matrices, $\alpha_i$ are combined orbital and spin indices, $u_{\vb{k} n s}$ are the normalized band eigenvectors which diagonalize the one-particle Hamiltonian, $H(\vb{k}) u_{\vb{k} n s} = \varepsilon_{\vb{k} n} u_{\vb{k} n s}$, and $\Theta$ is the unitary matrix that determines how single-particle states transform under the antiunitary time-reversal operator, $\hat{\Theta}^{-1} \psi_{\alpha_1}(\vb{k}) \hat{\Theta} = \sum_{\alpha_2} \Theta_{\alpha_1 \alpha_2}^{*} \psi_{\alpha_2}(-\vb{k})$.
A pseudospins degeneracy requires both time- and space-inversion symmetry, which we henceforth assume.

For the plasmon-mediated monopole and dipole interaction of Eq.~\eqref{eq:Hint-vq}, the monopole and dipole fermionic bilinears of Eq.~\eqref{eq:dipole_real_space} we write in the following form 
\begin{align}
D_{\mu}(\vb{q}) &= - e \sum_{\vb{k}} \psi^{\dag}(\vb{k}) \Gamma_{\mu}(\vb{k}, \vb{k}+\vb{q}) \psi(\vb{k}+\vb{q}). \label{eq:Pmu-assumed-form}
\end{align}
The pairing interaction then reads:
\begin{align}
W_{ba}(\vb{p}_m, \vb{k}_n) &= - \frac{M_{ba}(\vb{p}_m, \vb{k}_n) + M_{ba}(\vb{p}_m, - \vb{k}_n) p_a}{4 \abs{\grad_{\vb{p}} \varepsilon_{\vb{p} m}}^{1/2} \abs{\grad_{\vb{k}} \varepsilon_{\vb{k} n}}^{1/2}}, \label{eq:W-pairing-def}
\end{align}
where $p_{a=0} = - p_{a=1,2,3} = +1$ and
\begin{align}
\begin{aligned}
M_{ba}(\vb{p}_m, \vb{k}_n) &= \sum_{\mu \nu} e^2 V_{\mu \nu}(\vb{p}-\vb{k}) \\[-6pt]
&\hspace{24pt} \times \Tr P_{\vb{p} m}^{b} \Gamma_{\mu}(\vb{p}, \vb{k}) P_{\vb{k} n}^{a} \Gamma_{\nu}^{\dag}(\vb{p}, \vb{k}).
\end{aligned}
\end{align}
Here we used the fact that $D_{\mu}$ are even under time-reversal, $\hat{\Theta}^{-1} D_{\mu}(\vb{q}) \hat{\Theta} = D_{\mu}(-\vb{q})$. $D_{\mu}$ are also Hermitian, $D_{\mu}^{\dag}(\vb{q}) = D_{\mu}(-\vb{q})$, so that $\Gamma_{\nu}^{\dag}(\vb{p}, \vb{k}) = \Gamma_{\nu}(\vb{k}, \vb{p})$.
The trace arising in $M_{ba}(\vb{p}_m, \vb{k}_n)$ goes over spin and orbital degrees of freedom and one can alternatively write it as a pseudospin trace:
\begin{align}
\begin{aligned}
&\Tr P_{\vb{p} m}^{b} \Gamma_{\mu}(\vb{p}, \vb{k}) P_{\vb{k} n}^{a} \Gamma_{\nu}^{\dag}(\vb{p}, \vb{k}) = \\
&\hspace{50pt} = \tr \Pauli_b \tilde{\gamma}_{\mu}(\vb{p}_m, \vb{k}_n) \Pauli_a \tilde{\gamma}_{\nu}^{\dag}(\vb{p}_m, \vb{k}_n),
\end{aligned}
\end{align}
where
\begin{align}
\mleft[\tilde{\gamma}_{\mu}(\vb{p}_m, \vb{k}_n)\mright]_{s's} &\defeq u_{\vb{p} m s'}^{\dag} \Gamma_{\mu}(\vb{p}, \vb{k}) u_{\vb{k} n s}.
\end{align}

\subsection{General properties and estimates for pairing mediated by electric monopole-dipole interactions} \label{sec:qualitative-pairing}
The fact that all interactions between the electric monopoles and dipoles are mediated by the same electrostatic field allows us to make a number of very general statements regarding the pairing.
We start by writing
\begin{align}
V_{\mu \nu}(\vb{q}) &= v_{\mu}(\vb{q}) V(\vb{q}) v_{\nu}^{*}(\vb{q}), \label{eq:vVv-decomp}
\end{align}
where $v_{\mu}(\vb{q}) = \mleft(1, \iu q_i\mright)$, to encode this fact; see also Fig.~\ref{fig:plasmon-fermion-vertex}.
After renormalization, only $V(\vb{q}) \to \mathscr{V}(\vb{q})$ changes.
It then follows that
\begin{align}
M_{00}(\vb{p}_m, \vb{k}_n) &= e^2 V(\vb{p}-\vb{k}) \sum_{s's} \abs{\mleft[\overline{\gamma}(\vb{p}_m, \vb{k}_n)\mright]_{s's}}^2
\end{align}
is strictly positive in the singlet channel, with $\overline{\gamma}$ given by
\begin{align}
\mleft[\overline{\gamma}(\vb{p}_m, \vb{k}_n)\mright]_{s's} &= \sum_{\mu=0}^{3} v_{\mu}(\vb{p}-\vb{k}) \mleft[\tilde{\gamma}_{\mu}(\vb{p}_m, \vb{k}_n)\mright]_{s's}.
\end{align}
The singlet pairing interaction $W_{00}(\vb{p}_m, \vb{k}_n)$ is thus negative-definite.
For negative-definite matrices, the Perron-Frobenius theorem~\cite{Berman1994} applies and states that the largest-in-magnitude eigenvalue $\lambda_{\star}$ is negative and that the corresponding eigenvector $d_{\star}(\vb{k}_n)$ has no nodes, i.e., is $s$-wave.
While $\lambda_{\star}$ and $d_{\star}(\vb{k}_n)$ do not correspond to a superconducting instability, they are nonetheless a useful reference that bounds the possible pairing instabilities.
In particular, all positive singlet eigenvalues are bounded by $\abs{\lambda_{\star}}$ and to be orthogonal to $d_{\star}(\vb{k}_n)$ their eigenvectors need to either have nodes or sign changes between Fermi surfaces.
Hence any singlet superconductivity must be unconventional and weaker than $\abs{\lambda_{\star}}$.

The triplet eigenvalues are bounded by $\abs{\lambda_{\star}}$ as well.
To show this, consider the eigenvector corresponding to the largest triplet eigenvalue.
Using the $\SU(2)$ pseudospin gauge freedom, we may always orient this eigenvector along the $\vu{e}_3 = \vu{z}$ direction.
The corresponding
\begin{align}
\begin{aligned}
M_{33}(\vb{p}_m, \vb{k}_n) &= e^2 V(\vb{p}-\vb{k}) \\
&\hspace{16pt} \times \sum_{s's} (\pm)_{s'} (\pm)_{s} \abs{\mleft[\overline{\gamma}(\vb{p}_m, \vb{k}_n)\mright]_{s's}}^2
\end{aligned}
\end{align}
is therefore bounded by $M_{00}(\vb{p}_m, \vb{k}_n)$, as is $W_{33}(\vb{p}_m, \vb{k}_n)$ by $\abs{W_{00}(\vb{p}_m, \vb{k}_n)}$.
A corollary of the Perron-Frobenius theorem~\cite{Berman1994} then states that the largest-in-magnitude triplet eigenvalue is smaller in magnitude than the largest-in-magnitude singlet eigenvalue, which we wanted to show.
That said, the largest \emph{positive} triplet eigenvalue may still be larger than the largest \emph{positive} singlet eigenvalue, resulting in triplet pairing overall.
Compare with similar statements for electron-phonon mediated superconductivity~\cite{Brydon2014}.

Although it is, of course, expected that electronic mechanisms can only give superconductivity that is unconventional (not $s$-wave), the arguments of the previous paragraphs show this rigorously.
More interesting is the statement that the Cooper pairing strength is bounded by the strength of the repulsion, as measured by $\lambda_{\star}$.
To get an intuition regarding $\lambda_{\star}$, let us consider the simplest limit where there is only monopole coupling with $\Gamma_{\mu=0}(\vb{k}, \vb{k}+\vb{q}) = \one$ in Eq.~\eqref{eq:Pmu-assumed-form}.
We may then schematically write
\begin{align}
\begin{aligned}
\lambda_{\star} &\approx - \frac{1}{2} g_F \ev{\frac{e^2}{\epsilon_0 \vb{q}^2 + e^2 g_F}}_{\text{FS}} \\
&\sim - \frac{1}{2} \, \frac{e^2 g_F}{\epsilon_0 k_F^2} \log\mleft(1 + \frac{\epsilon_0 k_F^2}{e^2 g_F}\mright),
\end{aligned} \label{eq:schem-lambdaStar}
\end{align}
where in the interaction $V(\vb{q})$ we included Thomas-Fermi screening, $k_F$ characterizes the size of the Fermi sea ($\sim k_F^2$ is the area), and the total density of states (DOS) is
\begin{align}
g_F &= 2 \sum_n \int_{\varepsilon_{\vb{k} n} = 0} \frac{\dd{S_{\vb{k}}}}{(2\pi)^3} \frac{1}{\abs{\grad_{\vb{k}} \varepsilon_{\vb{k} n}}}.
\end{align}

Hence $\lambda_{\star}$ goes like $\sim g_F \abs{\log g_F}$ to zero for small $g_F$, and to $-1/2$ for large $g_F$.
Clearly then, a small DOS is unfavorable for superconductivity, as expected.
Less obviously, one cannot make the pairing arbitrarily strong by increasing the DOS because of the DOS-dependent screening.
This is in distinction to other mechanisms, such as pairing due to phonons and, to some extent, also the pairing due to quantum-critical boson exchange~\cite{Millis2001, Chubukov2003, Lederer2015, Lederer2017, Klein2018, Klein2023}, where the DOS can be increased while the pairing interaction changes only moderately.

\begin{figure}[t]
\includegraphics[width=0.95\columnwidth]{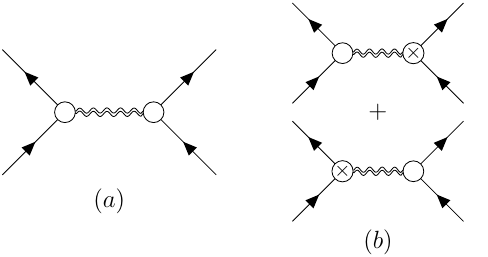}
\caption{The leading contributions to the pairing interaction derive from monopole-monopole (a) and monopole-dipole (b) coupling.
The double wiggly lines indicates that the plasmon propagator is screened.
With sufficient screening, the repulsive contribution from (a) mainly acts in the $s$-wave channel and is orthogonal to the attractive contribution from (b).}
\label{fig:monopole-dipole-exchange}
\end{figure}

Finally, we show that our interaction can indeed have positive eigenvalues, resulting in superconductivity, when the screening and dipole moments are strong enough.
Our reasoning is the following:
In the interaction~\eqref{eq:vVv-decomp}, $V(\vb{q})$ decreases with increasing $\vb{q}$, whereas the dipolar part of $v_{\mu}(\vb{q}) = \mleft(1, \iu q_i\mright)$ increases.
Strong screening means that $V(\vb{q})$ decays weakly with increasing $\vb{q}$.
Thus sufficiently large electric dipole moments can overwhelm this decay to give an interaction that is overall more strongly repulsive at finite $\vb{q}$ than at $\vb{q} = \vb{0}$.
It then follows that pairing eigenvectors which change sign every $\vb{Q}$, where $\vb{Q} \neq \vb{0}$ is the repulsion peak, have positive eigenvalues~\cite{Maiti2013}, which we wanted to show.
A similar behavior occurs in the celebrated Kohn-Luttinger mechanism~\cite{Kohn1965, Luttinger1966, Maiti2013, Kagan2014} in which the overscreening of $V(\vb{q})$ is a consequence of the $2 k_F$ non-analyticity of the system.
In our case, the electric dipoles are responsible for this overscreening and formally it develops already in the leading order of the Coulomb interaction (Fig.~\ref{fig:monopole-dipole-exchange}).
In particular, to leading order in powers of the electric dipole moment, the interaction that is responsible for the pairing in our mechanism is the screened monopole-dipole interaction shown in Fig.~\ref{fig:monopole-dipole-exchange}(b).

To illustrate our mechanism, we consider a Fermi liquid with spherical symmetry and only one Fermi surface.
For the interaction and coupling we assume
\begin{align}
\begin{aligned}
e^2 V(\vb{p}-\vb{k}) &= U_0 + U_1 \vu{p} \vdot \vu{k} + \cdots, \\
\tilde{\gamma}_{0}(\vb{p}, \vb{k}) &= \Pauli_0, \\
\tilde{\gamma}_{i}(\vb{p}, \vb{k}) &= - \frac{\eta}{2 e} \mleft[(\vu{p} + \vu{k}) \vcross \vb{\Pauli}\mright]_{i},
\end{aligned}
\end{align}
where $\vu{p} = \vb{p} / \abs{\vb{p}}$, $\vu{k} = \vb{k} / \abs{\vb{k}}$, and $\abs{\vb{p}} = \abs{\vb{k}} = k_F$.
$U_1 > 0$ quantifies the degree of screening and $\eta$ is the electric dipole moment.
To linear order in $U_1$ and $\eta$:
\begin{gather}
\begin{gathered}
W_{00}(\vb{p}, \vb{k}) = - \frac{U_0}{v_F}, \\
W_{ij}(\vb{p}, \vb{k}) = 2 \frac{U_0 k_F \eta}{v_F e} (\hat{p}_i \hat{k}_j - \hat{p}_j \hat{k}_i) - \frac{U_1}{v_F} \vu{p} \vdot \vu{k} \, \Kd_{ij},
\end{gathered}
\end{gather}
where $i, j \in \{1, 2, 3\}$ and $v_F = \abs{\grad_{\vb{k}} \varepsilon_{\vb{k}}}$ is the Fermi velocity at $\abs{\vb{k}} = k_F$.
In the singlet channel we find no pairing, while for the leading instability in the triplet channel we find
\begin{gather}
\begin{gathered}
\lambda = \frac{2}{3} g_F U_0 k_F \eta / e - \frac{1}{6} g_F U_1, \\
\vb{d}(\vb{k}) = \vu{k}
\end{gathered}
\end{gather}
which has pseudoscalar symmetry ($\sim \vu{k} \vdot \vb{\Pauli}$).
There is also a subleading $p$-wave instability with
\begin{gather}
\begin{gathered}
\lambda' = \frac{1}{3} g_F U_0 k_F \eta / e - \frac{1}{6} g_F U_1, \\
\vb{d}_{i}'(\vb{k}) = \vu{e}_{i} \vcross \vu{k}
\end{gathered}
\end{gather}
which is threefold degenerate; $i \in \{1, 2, 3\}$ and $\vu{e}_i$ are Cartesian unit vectors.
Thus if dipole moments are strong compared to the screening, namely $k_F \eta /e > U_1 / (4 U_0)$, the monopole-dipole electrostatic interaction will result in superconductivity.

\section{Cooper pairing in quasi-2D \\ Dirac metals} \label{sec:Dirac-pairing}
\begin{figure}[t]
\includegraphics[width=\columnwidth]{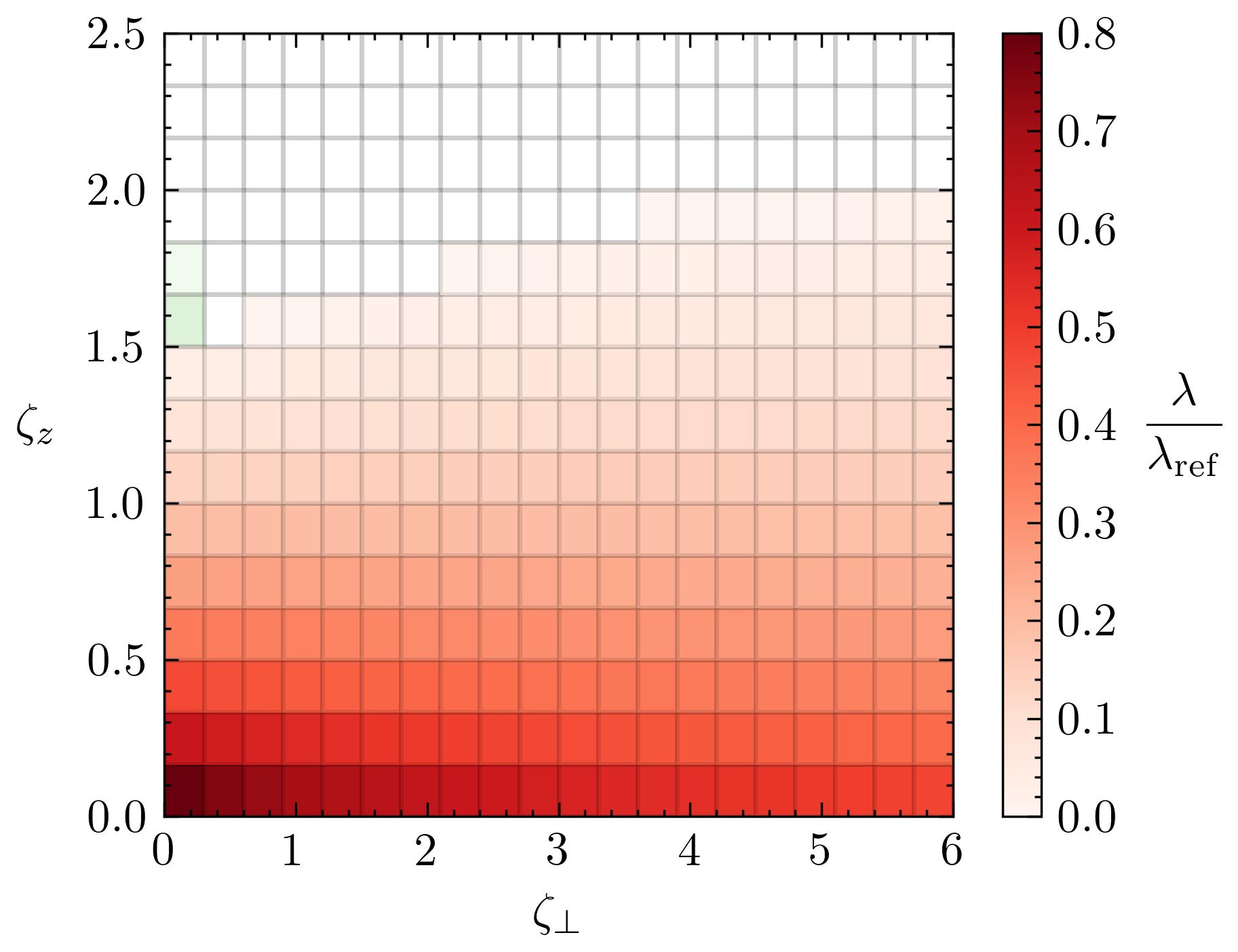} 
\caption{The largest pairing eigenvalue $\lambda$ as a function of the screening parameters $\zeta_{\perp}$ and $\zeta_z$ entering the RPA interaction of Eq.~\eqref{eq:RPA-int} for the case $\hat{m} = 1$ with $\hat{\eta} = 0.3$ [Eq.~\eqref{eq:new-meta}].
$\lambda$ is found by numerically solving Eq.~\eqref{eq:lin-gap-eq} on a dense grid and the reference $\lambda_{\text{ref}} = \abs{\hat{\eta}} / 4 \sqrt{1 + \hat{m}^2} \approx 0.05$ is given in Eq.~\eqref{eq:leading-pairing}.
The leading pairing state has pseudoscalar ($A_{1u}$) symmetry in regions colored red, which is almost everywhere.
Colored white are the regions of large $\zeta_z$ where there is no pairing.
On the two points around $(\zeta_{\perp}, \zeta_{z}) \approx (0, 1.7)$ highlighted green, the leading pairing state is $p$-wave with a small $\lambda / \lambda_{\text{ref}} \approx 0.01$.}
\label{fig:pairing-w-screening}
\end{figure}

Here we study the superconducting instabilities of the dipolar Dirac model of Sec.~\ref{sec:Dirac-model} in the quasi-2D limit $v_z \approx 0$, that is $v_z \Lambda_z \ll m$.
The starting point our analysis is the effective model that emerges at the end of the RG flow of Sec.~\ref{sec:Dirac-RG}.
This effective model has a negligible in-plane dipole coupling $\eta_{\perp} \approx 0$, an enhanced out-of-plane dipole coupling $\eta_z$, and a momentum cutoff $\Lambda \sim k_F$.
Its Cooper pairing we analyze using the linearized gap equation we introduced in Sec.~\ref{sec:lin-gap-eq}.
For strong enough screening and $z$-axis dipole moments $\eta_z$, we find that unconventional odd-parity Cooper pairing takes place which has pseudoscalar symmetry $\sim \vb{k} \vdot \vb{\Pauli}$, similar to the superfluid state of \ce{^3He-B}; see Fig.~\ref{fig:pairing-w-screening}.
In addition, we find a competitive subleading pairing instability of $p$-wave symmetry.

As in our RG treatment, we employ a large-$N$ expansion to analytically access the regime of strong screening.
A slight difference from Sec.~\ref{sec:Dirac-RG} is that the cutoff is not imposed on the frequencies [Eq.~\eqref{eq:cutoff-def}], but only on the momenta through their energies $\varepsilon_{\vb{k}} = \sqrt{m^2 + v^2 \vb{k}_{\perp}^2} - \mu$.
Because we ended the RG flow with a $\Lambda \sim k_F$, our energy cutoff $\hbar \omega_c$ is on the order of the Fermi energy $E_F = \mu - m = \sqrt{m^2 + v^2 k_F^2} - m$.
Note that the same convention with the energy cutoff was used in the derivation of Eqs.~\eqref{eq:lin-gap-eq} and~\eqref{eq:BCS-Tc-expr}.

Another minor difference from before is that we need to impose periodicity along the $\vu{z}$ direction on the model.
Instead of Eqs.~\eqref{eq:V-1-phi} and~\eqref{eq:A-psi}, we thus use
\begin{align}
V^{-1}(q) &= \epsilon_{\perp} \vb{q}_{\perp}^2 + \frac{4 \Lambda_z^2 \epsilon_z}{\pi^2} \sin^2\frac{\pi q_z}{2 \Lambda_z}, \\
A(k, p) &= e \gamma_0 + \iu \frac{\Lambda_z \eta_z}{\pi} \sin\frac{\pi (k_z - p_z)}{\Lambda_z} \gamma_0 \gamma_3.
\end{align}
This is necessary because we are interested in momenta with $\abs{\vb{q}_{\perp}} \sim k_F$ and $q_z \sim \Lambda_z$; cf.\ remarks after Eq.~\eqref{eq:VUmklapp}.
We only consider quasi-2D systems with $v_z = \eta_{\perp} = 0$ because of the RG considerations discussed before Eq.~\eqref{eq:quasi-2D-param}.

In the limit of strong screening, the interaction is given by the polarization bubble which in the static $\omega_q = 0$ limit for $\abs{\vb{q}_{\perp}} \leq 2 k_F$ equals (Appendix~\ref{sec:polarization}):
\begin{align}
\Pi(\omega_q = 0, \vb{q}) &= N g_F \mleft[e^2 + \frac{\Lambda_z^2 \eta_z^2}{\pi^2} \sin^2\frac{\pi q_z}{\Lambda_z}\mright], \label{eq:static-smallq-pol}
\end{align}
where
\begin{align}
g_F &= \frac{\Lambda_z \mu}{\pi^2 v^2}, &
\mu &= \sqrt{m^2 + v^2 k_F^2}.
\end{align}
Although this was evaluated without a cutoff ($\Lambda \to \infty$), reintroducing it does not significantly influence this expression.

To calculate the pairing interaction $W_{ab}(\vb{p}, \vb{k})$ of Eq.~\eqref{eq:W-pairing-def}, we need to diagonalize the Dirac Hamiltonian~\eqref{eq:Dirac-Haml}.
The dispersion of the conduction band is
\begin{align}
\varepsilon_{\vb{k}} &= \sqrt{m^2 + v^2 \vb{k}_{\perp}^2} - \mu,
\end{align}
and the corresponding conduction band eigenvectors are
\begin{align}
u_{\vb{k}\uparrow} &= \frac{1}{\sqrt{\mathcal{N}_{\vb{k}}}} \begin{pmatrix}
m + \sqrt{m^2 + v^2 \vb{k}_{\perp}^2} \\ 0 \\ 0 \\ - v (k_x + \iu k_y)
\end{pmatrix}, \\
u_{\vb{k}\downarrow} &= \frac{1}{\sqrt{\mathcal{N}_{\vb{k}}}} \begin{pmatrix}
0 \\ m + \sqrt{m^2 + v^2 \vb{k}_{\perp}^2} \\ v (k_x - \iu k_y) \\ 0
\end{pmatrix},
\end{align}
where $\uparrow, \downarrow$ are pseudospins, $u_{\vb{k}\uparrow} = P \Theta u_{\vb{k}\downarrow}^{*} = \iu \Pauli_y u_{\vb{k}\downarrow}^{*}$, and
\begin{align}
\mathcal{N}_{\vb{k}} &= 2 \sqrt{m^2 + v^2 \vb{k}_{\perp}^2} \mleft(m + \sqrt{m^2 + v^2 \vb{k}_{\perp}^2}\mright).
\end{align}
Recall that bolded vectors with $\perp$ subscripts are in-plane [Eq.~\eqref{eq:bold-vec-conv}].
The in-plane momenta that are on the cylindrical Fermi surface we parameterize with azimuthal angles:
\begin{align}
\begin{aligned}
\vb{p}_{\perp} &= k_F (\cos \theta_p, \sin \theta_p), \\
\vb{k}_{\perp} &= k_F (\cos \theta_k, \sin \theta_k).
\end{aligned}
\end{align}
Now it is straightforward to find $W_{ab}(\theta_p, p_z, \theta_k, k_z)$ as given by Eq.~\eqref{eq:W-pairing-def}.
The final expression for $W_{ab}$ that one obtains is fairly complicated, and one cannot diagonalize it [Eq.~\eqref{eq:lin-gap-eq}] analytically for general momentum-dependent interactions $\mathscr{V}(q)$.
Thus one needs to resort to numerical methods.

Physically, we are interested in the limit of strong screening in which case the momentum dependence of $\mathscr{V}(q)$ is weak.
To understand this limit, a good starting point is to consider a constant Hubbard-like interaction
\begin{align}
\mathscr{V}(q) &= \frac{1}{g_F e^2} \equiv U_0 \label{eq:perfect-screening}
\end{align}
which corresponds to the large-$N$ limit [Eq.~\eqref{eq:static-smallq-pol}] with the $q_z$ dependence neglected.
The numerical results which we will shortly present can be well understood by analyzing this idealized scenario. 
For a constant interaction, we can exactly diagonalize $W_{ab}$.
The result is:
\begin{align}
W(p, k) &= \frac{e^2 U_0}{v} \sum_{n=1}^{12} w_n \sum_{s=1}^{\dim n} d_{n,s}(p) d_{n,s}^{\dag}(k), \label{eq:W-eigs-def}
\end{align}
where $w_n$ are dimensionless eigenvalues of degeneracy $\dim n$ and $d_{n,s}(p) = d_{n,s}(\theta_p, p_z)$ are the corresponding eigenvectors.
Both are listed in Table~\ref{tab:W-eigs}.
Here we used the shorthand
\begin{align}
p &\defeq (\theta_p, p_z), &
k &\defeq (\theta_k, k_z).
\end{align}
The eigenvectors are orthogonal and normalized according to
\begin{align}
\int_{-\pi}^{\pi} \frac{\dd{\theta_k}}{2 \pi} \int_{-\Lambda_z}^{\Lambda_z} \frac{\dd{k_z}}{2 \pi} \, d_{n,s}^{\dag}(k) d_{n',s'}(k) &= \frac{\Lambda_z}{\pi} \Kd_{nn'} \Kd_{ss'}.
\end{align}
The corresponding pairing eigenvalues $\lambda$ arising in the linearized gap equation~\eqref{eq:lin-gap-eq} therefore equal
\begin{align}
\lambda_n &= \frac{w_n}{2 \sqrt{1 + \hat{m}^2}},
\end{align}
where
\begin{align}
\hat{m} &\defeq \frac{m}{v k_F}, &
\hat{\eta} &\defeq \frac{\Lambda_z \eta_z}{\pi \, e} \label{eq:new-meta}
\end{align}
are dimensionless measures of the gap and electric dipole coupling, respectively.
Given how $\Lambda_z / \pi$ arises in many places, we shall find it convenient to henceforth set $\Lambda_z = \pi$, that is, the lattice constant along $z$ to unity.

\begin{table*}
\caption{The eigenvalues $w_n$ and eigenvectors $d_{n,s}(k) = d_{n,s}(\theta_k, k_z)$ arising in the expression~\eqref{eq:W-eigs-def} for the pairing interaction $W_{ab}(p, k)$ of a quasi-2D Dirac metal with a constant interaction.
Here $\hat{m} \defeq m / (v k_F)$, $\hat{\eta} \defeq \Lambda_z \eta_z / (\pi \, e)$, $\Lambda_z = \pi$, and $\theta_k$ is the azimuthal angle, $\vb{k}_{\perp} = k_F (\cos \theta_k, \sin \theta_k)$, specifying the in-plane position on the cylindrical Fermi surface.
The degeneracy $\dim n$ of the $n$-th eigenvalue is either $1$ or $2$, depending on how many eigenvectors are shown in the table.
For even-parity pseudospin-singlet eigenvectors only the first component is finite, while in odd-parity pseudospin-triplet eigenvectors only the last three components are finite and together constitute the corresponding Balian-Werthamer $\vb{d}$-vector.}
{\renewcommand{\arraystretch}{2.3}
\renewcommand{\tabcolsep}{11.5pt}
\begin{tabular}{lcc} \hline\hline
$n$ & $w_n$ & $d_{n,s}(\theta_k, k_z)$ \\ \hline
$1$ & $\displaystyle - \frac{1 + 2 \hat{m}^2 + \hat{\eta}^2}{2 \sqrt{1+\hat{m}^2}}$ & $\mleft(1,~ 0,~ 0,~ 0\mright)$ \\
$2$ & $\displaystyle - \frac{1 - \hat{\eta}^2}{2 \sqrt{1+\hat{m}^2}}$ & $\mleft(0,~ \cos \theta_k,~ \sin \theta_k,~ 0\mright)$ \\
$3$ & $\displaystyle - \frac{1 + \hat{\eta}^2}{2 \sqrt{1+\hat{m}^2}}$ & $\mleft(0,~ - \sin \theta_k,~ \cos \theta_k,~ 0\mright)$ \\
$4$ & $\displaystyle - \frac{2 - \hat{\eta}^2}{4 \sqrt{1+\hat{m}^2}}$ & $\begin{cases}
\mleft(0,~ 0,~ 0,~ \sqrt{2} \cos \theta_k\mright) \\
\mleft(0,~ 0,~ 0,~ \sqrt{2} \sin \theta_k\mright)
\end{cases}$ \\
$5$ & $\displaystyle \frac{\hat{\eta}}{2}$ & $\mleft(0,~ \cos \theta_k \cos k_z,~ \sin \theta_k \cos k_z,~ \sin k_z\mright)$ \\
$6$ & $\displaystyle - \frac{\hat{\eta}}{2}$ & $d_{5}(\theta_k, -k_z)$ \\
$7$ & $\displaystyle \frac{\hat{\eta} \sqrt{1 + 2 \hat{m}^2}}{4 \sqrt{1+\hat{m}^2}}$ & $\begin{cases}
\mleft(0,~ \frac{\sqrt{1 + \hat{m}^2} + \hat{m} + \mleft(\sqrt{1 + \hat{m}^2} - \hat{m}\mright) \cos 2 \theta_k}{\sqrt{2 (1 + 2 \hat{m}^2)}} \sin k_z,~ \frac{\mleft(\sqrt{1 + \hat{m}^2} - \hat{m}\mright) \sin 2 \theta_k}{\sqrt{2 (1 + 2 \hat{m}^2)}} \sin k_z,~ - \sqrt{2} \cos \theta_k \cos k_z\mright) \\
\mleft(0,~ \frac{\mleft(\sqrt{1 + \hat{m}^2} - \hat{m}\mright) \sin 2 \theta_k}{\sqrt{2 (1 + 2 \hat{m}^2)}} \sin k_z,~ \frac{\sqrt{1 + \hat{m}^2} + \hat{m} - \mleft(\sqrt{1 + \hat{m}^2} - \hat{m}\mright) \cos 2 \theta_k}{\sqrt{2 (1 + 2 \hat{m}^2)}} \sin k_z,~ - \sqrt{2} \sin \theta_k \cos k_z\mright)
\end{cases}$ \\
$8$ & $\displaystyle - \frac{\hat{\eta} \sqrt{1 + 2 \hat{m}^2}}{4 \sqrt{1+\hat{m}^2}}$ & $\begin{cases}
d_{7,1}(\theta_k, -k_z) \\
d_{7,2}(\theta_k, -k_z)
\end{cases}$ \\
$9$ & $\displaystyle \frac{\hat{\eta}^2}{4\sqrt{1+\hat{m}^2}}$ & $\sqrt{2} \cos(2 k_z) \times d_{1}(\theta_k, k_z)$ \\
$10$ & $\displaystyle - \frac{\hat{\eta}^2}{4\sqrt{1+\hat{m}^2}}$ & $\sqrt{2} \cos(2 k_z) \times d_{2}(\theta_k, k_z)$ \\
$11$ & $\displaystyle \frac{\hat{\eta}^2}{4\sqrt{1+\hat{m}^2}}$ & $\sqrt{2} \cos(2 k_z) \times d_{3}(\theta_k, k_z)$ \\
$12$ & $\displaystyle - \frac{\hat{\eta}^2}{8\sqrt{1+\hat{m}^2}}$ & $\begin{cases}
\sqrt{2} \cos(2 k_z) \times d_{4,1}(\theta_k, k_z) \\
\sqrt{2} \cos(2 k_z) \times d_{4,2}(\theta_k, k_z)
\end{cases}$
\\ \hline\hline
\end{tabular}}
\label{tab:W-eigs}
\end{table*}

Of the twelve $w_n$, four are positive and give positive $\lambda$ which correspond to superconducting instabilities.
The leading instability among these four is odd-parity and pseudospin-triplet, with ($n = 5, 6$ in Table~\ref{tab:W-eigs})
\begin{gather}
\begin{gathered}
\lambda = \frac{\abs{\hat{\eta}}}{4 \sqrt{1 + \hat{m}^2}}, \\
\vb{d}(k) = \begin{pmatrix}
\cos \theta_k \cos k_z \\
\sin \theta_k \cos k_z \\
\sgn \hat{\eta} \, \sin k_z
\end{pmatrix}.
\end{gathered} \label{eq:leading-pairing}
\end{gather}
Since $\vb{d}(k) \sim (k_x, k_y, \pm k_z)$, its symmetry is pseudoscalar.
The subleading pairing instability is also odd-parity and pseudospin-triplet, but has $p$-wave symmetry and is weaker by a factor in between $\sqrt{2}$ and $2$ from the leading instability.
It is a two-component pairing state that may either give rise to time-reversal symmetry breaking or nematic superconductivity, depending on the quartic coefficients in the Ginzburg-Landau expansion.
Its pairing eigenvalue equals:
\begin{gather}
\lambda' = \frac{\abs{\hat{\eta}} \sqrt{1 + 2 \hat{m}^2}}{8 (1 + \hat{m}^2)}. \label{eq:subleading-pairing}
\end{gather}
The corresponding two degenerate eigenvectors are fairly complicated and are provided in the $n = 7, 8$ entries of Table~\ref{tab:W-eigs}.
In agreement with our general discussion of Sec.~\ref{sec:qualitative-pairing}, the largest-in-magnitude $\lambda$ which bounds all other $\lambda$ is ($n = 1$ in Table~\ref{tab:W-eigs})
\begin{align}
\lambda_{\star} &= - \frac{1 + 2 \hat{m}^2 + \hat{\eta}^2}{4 (1 + \hat{m}^2)}
\end{align}
and it has an even-parity pseudospin-singlet $s$-wave eigenvector; cf.\ Eq.~\eqref{eq:schem-lambdaStar}.

A more realistic screened interaction is given by RPA:
\begin{align}
\mathscr{V}(q) &= \frac{U_0}{1 + \zeta_{\perp} \sin^2\frac{\theta_q}{2} + \zeta_z \sin^2\frac{q_z}{2} + \hat{\eta}^2 \sin^2 q_z}, \label{eq:RPA-int}
\end{align}
where $q = (\theta_q, q_z) = p - k$, $U_0 \defeq 1 / (g_F e^2)$, and the strength of the screening we specify using the dimensionless parameters:
\begin{align}
\begin{aligned}
\zeta_{\perp} &\defeq \frac{4 k_F^2 \epsilon_{\perp}}{g_F e^2} = \frac{4 \pi}{\sqrt{1 + \hat{m}^2}} \frac{\pi k_F}{\Lambda_z} \frac{v \epsilon_{\perp}}{e^2}, \\
\zeta_z &\defeq \frac{4 \Lambda_z^2 \epsilon_z}{\pi^2 g_F e^2} = \frac{4 \pi}{\sqrt{1 + \hat{m}^2}} \frac{\Lambda_z}{\pi k_F} \frac{v \epsilon_z}{e^2}.
\end{aligned}
\end{align}
For such a $\mathscr{V}(q)$, we have numerically investigated the resulting pairing instabilities.
The results for one generic parameter choice are shown in Fig.~\ref{fig:pairing-w-screening}.
For general parameter sets, we find that pairing takes place only when $\zeta_{\perp}$ and $\zeta_z$ are sufficiently small compared to $\abs{\hat{\eta}}$.
This agrees with the conclusions drawn from the schematic example we considered at the end of Sec.~\ref{sec:qualitative-pairing}.
Moreover, the symmetry of the leading pairing state is robustly pseudoscalar triplet, with essentially the same $\vb{d}$-vector as in Eq.~\eqref{eq:leading-pairing}.
A $p$-wave instability also arises that, although usually weaker by a factor of $\sim \sqrt{2}$ than the leading instability, in a few cases becomes leading.

In many materials, $v$ is on the order of \SI{1}{\electronvolt\angstrom} which gives a small $\alpha^{-1} = v \epsilon_0 / e^2 \sim 0.006$. 
Hence for $\hat{m} \sim 1$, $k_F / \Lambda_z \sim 1$, and $\epsilon_{\perp} \sim \epsilon_z \sim \epsilon_0$ the screening coefficients $\zeta_{\perp}$ and $\zeta_z$ can be very small, i.e., the screening can be very efficient.
In other words, for physically realistic parameters the momentum-dependence of the screened interaction can be such that it only modestly suppresses the pairing eigenvalue $\lambda$ from its $\zeta_{\perp} = \zeta_z = 0$ value~\eqref{eq:leading-pairing}.
That said, one should keep in mind that $\alpha$ flows toward weak coupling under RG, as shown in Sec.~\ref{sec:Dirac-RG}; see Fig.~\ref{fig:RG-alpha-flow}.
For the materials that motivated the current study, one finds $v \sim \SI{3}{\electronvolt\angstrom}$ in the case of \ce{Bi2Se3} and \ce{Bi2Te3}~\cite{Liu2010} and $v \sim \SI{1}{\electronvolt\angstrom}$ in the case of \ce{SnTe}~\cite{Hsieh2012}.
The dielectric constants are up to $\sim 10$ in the frequency range of interest for these materials~\cite{Eddrief2016, Lewis1987, Suzuki1995}, giving a small enough $\alpha^{-1} \sim 0.2$ for our theory to be of relevance.

We have thus found that the leading paring instability is odd-parity and of pseudoscalar ($A_{1u}$) symmetry.
It is interesting to note that states of such symmetry are more robust to disorder than usual~\cite{Dentelski2020}.
As demonstrated in Ref.~\cite{Dentelski2020}, this follows from the fact that the pseudoscalar pairing state transforms like a singlet under the combined application of chiral and time-reversal symmetry, which in turn implies that it is protected by an effective Anderson theorem relative to disorder which respect these symmetries.

If we use $\hat{\eta} \approx 0.5$ as the largest value for the effective dimensionless electric-dipole coupling that follows from the RG treatment of Sec.~\ref{sec:Dirac-RG}, we obtain from Eq.~\eqref{eq:leading-pairing} a dimensionless pairing eigenvalue $\lambda \approx 0.1$ which puts the system into the weak-coupling BCS regime.
A quantitative estimate of the transition temperature requires knowledge of the cutoff energy $\hbar \omega_c$.
Using for example $E_F \sim \SI{1}{\electronvolt}$, which is the appropriate energy scale for an electronic mechanism, one gets transition temperatures in the sub-Kelvin regime.
While these transition temperatures are not large, they do give rise to unconventional pairing in materials without strong local electron correlations or quantum-critical fluctuations of any kind.

Interestingly, the leading pairing state~\eqref{eq:leading-pairing} can be interpreted as the quasi-2D solid-state analog of the B phase of superfluid \ce{^3He}~\cite{Leggett1975, Vollhardt1990, Volovik2003}.
In the helium case, it is known that this phase is topological in three dimensions~\cite{Volovik2003, Volovik2009, Mizushima2015}, belonging to the class DIII in the classification of non-interacting gapped topological matter~\cite{Chiu2016, Ludwig2016}.
Hence, it couples to gravitational instantons through a topological $\theta$ term and its boundary contains a Majorana cone of topologically-protected surface Andreev bound states~\cite{Mizushima2015, Ludwig2016}.
To test whether our state is topological, we have evaluated the corresponding topological invariant~\cite{Volovik2009, Mizushima2015} and found that it vanishes.
Hence our pairing state is topologically trivial.
As shown in Ref.~\cite{Fu2010}, fully-gapped odd-parity pairing states need to have a Fermi surface which encloses an odd number of time-reversal invariant momenta to be topological.
In our case, the cylindrical Fermi surface encloses not only the $\Gamma$ point, but also the $Z$ point $\vb{k} = (0, 0, \pi)$, unlike \ce{^3He-B}, which explains the difference in topology.

\section{Conclusion}
In this work, we developed the theory of electric dipole excitations of electronic states residing near the Fermi level, we demonstrated that out-of-plane electric dipole fluctuations become enhanced at low energies in spin-orbit-coupled quasi-2D Dirac systems, and we showed that electric monopole-dipole interactions induce unconventional low-temperature superconductivity in sufficiently screened systems.
In quasi-2D Dirac metals in particular, we found that the resulting pairing state is an odd-parity state of pseudoscalar ($A_{1u}$) symmetry, similar to the superfluid phase of \ce{^3He-B}~\cite{Leggett1975, Vollhardt1990, Volovik2003}, with a competitive subleading $p$-wave instability appearing as well.
These are the main results of our work.

In our general treatment of dipole fluctuations, we made two key observations.
The first one is that intraband electric dipole excitations require spin-orbit coupling to maintain a finite coupling to plasmons in the long-wavelength limit.
The second one is that the same plasmon field mediates all effective electric multipole-multipole interactions that arise from the electron-electron Coulomb interaction.
With these in mind, we then formulated a general theory of itinerant dipole excitations and their electrostatic interactions.
In addition, we related our treatment of dynamically fluctuating dipoles to the Modern Theory of Polarization~\cite{Resta1994, Resta2000} and showed that the King-Smith--Vanderbilt formula~\cite{KingSmith1993} for the (static) polarization acquires an anomalous term within tight-binding descriptions.

When strong spin-orbit coupling inverts bands of opposite parities, dipole fluctuations are especially strong.
The vicinity of such band-inverted points is, moreover, generically described by Dirac models.
Although this has been known in various particular cases~\cite{Cohen1960, Wolff1964, Rogers1968, Zhang2009, Liu2010}, in this paper we presented a general symmetry derivation of this important fact, before turning to the renormalization group analysis of dipole excitations in Dirac systems.
Our large-$N$ RG analysis of the strong-screening limit reveled that, although irrelevant in most systems, electric dipole coupling is marginally relevant along the out-of-plane direction in quasi-2D geometries.
Even though the enhancement of the effective $z$-axis (out-of-plane) dipole coupling is limited, it is sufficiently large to imply that electronic dipole interactions cannot be ignored at low energies.
As a concrete experimental footprint, we have found that this $z$-axis dipole coupling gives the dominant contribution to the $z$-axis optical conductivity in quasi-2D Dirac systems.

The electric monopole-dipole coupling between itinerant electrons, introduced in this work, causes unconventional superconductivity whenever dipole moments are sufficiently strong compared to the screening.
Even when other pairing mechanisms are present, as long as they mostly act in the $s$-wave channel which is suppressed by the electric monopole-monopole repulsion, electric monopole-dipole interactions can still be the main cause of pairing.
Hence, in systems not governed by strong local electronic correlations or nearly critical collective modes, the proposed mechanism is a possible source of unconventional low-temperature superconductivity.
Using arguments similar to those of Ref.~\cite{Brydon2014}, we showed that the pairing due to our mechanism is necessarily unconventional, but also that it is not likely to reach high-temperatures (strong-coupling).
For comparison, the pairing due to the exchange of phonons~\cite{Brydon2014}, ferroelectric modes~\cite{KoziiBiRuhman2019, Enderlein2020, Kozii2022, Volkov2022, Klein2023}, and non-magnetic odd-parity fluctuations~\cite{Kozii2015} robustly favors conventional $s$-wave pairing and is able to reach strong coupling.
A longer comparison of our theory to theories of ferroelectric metals~\cite{Benedek2016, Chandra2017, KoziiBiRuhman2019, Volkov2020, Enderlein2020, Kozii2022, Volkov2022, Klein2023} was made in the introduction.
Although we included dipole-dipole interactions in our analysis, we found that they give a weaker (subleading) contribution to the Cooper pairing for realistic dipole strengths.
This should be contrasted with pairing in degenerate dipolar Fermi gases~\cite{Baranov2002, Bruun2008, *Bruun2008-E, Sieberer2011, Liu2012, Baranov2012}, discussed in the introduction, in which the neutrality of the cold-atom fermions precludes monopole-dipole interactions, rendering dipole-dipole interactions dominant.

The pairing mechanism proposed in this work is similar to other electronic mechanism~\cite{Maiti2013, Kagan2014}, which derive in one form or another from the electron-electron Coulomb interaction.
In their pioneering study~\cite{Kohn1965, Luttinger1966}, Kohn and Luttinger showed that the non-analyticity originating from the sharpness of the Fermi surface induces pairing with high orbital angular momentum $\ell$ in isotropic 3D systems, even when the short-ranged bare interaction is repulsive in all channels.
Although non-analyticity has proven to be a negligible source of pairing, giving $T_c \sim \SI{e-11}{\kelvin}$ or smaller~\cite{Kohn1965}, the idea that the overscreening of a bare repulsive interaction can result in pairing has survived and been developed in many ways~\cite{Maiti2013, Kagan2014}.
Subsequent work generalized this mechanism to isotropic 2D systems~\cite{Chubukov1993} and low-density Hubbard models~\cite{Baranov1992, Baranov1992-p2, Kagan2011}, as well as showed that the pairing extends to $\ell = 1$ for a bare repulsive contact interaction in isotropic 3D systems~\cite{Fay1968, Kagan1988, Baranov1992-p2, Efremov2000}, with a $T_c \sim \SI{e-3}{\kelvin}$ when applied to \ce{^3He}~\cite{Kagan1988}.
For repulsive Hubbard models, asymptotically exact weak-coupling solutions were found which gave pairing in both $p$-wave and $d$-wave channels~\cite{Raghu2010, Raghu2011}.

In our mechanism, just like in the Kohn-Luttinger-like mechanisms, an initially repulsive interaction becomes overscreened, resulting in pairing.
Both mechanisms need the interaction to be, or become, nearly momentum-independent.
Because we had started from the long-ranged unscreened Coulomb interaction, to screen it properly we needed to reach the strong-coupling regime of large $\alpha = e^2 / (\hbar v_F \epsilon_0)$.
Since this regime cannot be analytically treated in the unmodified model~\cite{Chubukov1989, Raghu2012}, we employed a large-$N$ expansion, $N$ being the number of fermion flavors.
In contrast, Kohn-Luttinger-like mechanisms start from a short-ranged repulsive interaction which is readily perturbatively treated.
The origin of the overscreening is different between the two mechanisms as well.
In our mechanism, the electric dipole terms appearing in the bare vertex are responsible, and not perturbative corrections to the Cooper-channel interaction.
Once projected onto the Fermi surface, the dipolar part of the bare vertex acquires a non-trivial structure in pseudospin space which plays an important role in choosing the pairing symmetry.
In Kohn-Luttinger-like mechanisms, on the other hand, the pairing symmetry is essentially chosen by the momentum-dependence of the overscreened interaction.

In light of the strong dipole fluctuations we had found in quasi-2D Dirac systems, we explored their pairing instabilities.
Across most of the parameter range, the dominant pairing state due to electric monopole-dipole interactions has pseudoscalar ($A_{1u}$) symmetry and resembles the Balian-Werthamer state of \ce{^3He-B}~\cite{Leggett1975, Vollhardt1990, Volovik2003}.
Since the dimensionless dipole coupling is at best a fraction of the monopole coupling, the pairing problem is expected to be in the weak-coupling regime.
Although we estimated transition temperatures on order of \SI{0.1}{\kelvin}, a detailed prediction of $T_c$ will depend on a number of material parameters, making quantitative predictions rather unreliable.
That said, it is interesting to observe that \ce{SnTe} is well-described by Dirac models~\cite{Rogers1968, Hsieh2012, Ando2015} and that an $A_{1u}$ pairing state is consistent with experiments performed on \ce{In}-doped \ce{SnTe}~\cite{Novak2013, Zhong2013, Smylie2018-p2, Smylie2020, Saghir2014, Maeda2017}.
This suggests that our mechanism could be of relevance.
In the case of doped \ce{Bi2Se3}, which is also well-described by Dirac models~\cite{Zhang2009, Liu2010}, there is strong evidence for nematic $p$-wave pairing~\cite{Yonezawa2019, Matano2016, Pan2016, Yonezawa2017, Asaba2017, Du2017, Smylie2018, Cho2020, Yonezawa2017, Smylie2016, Smylie2017, Fu2010, Venderbos2016, Hecker2018}, which in our mechanism is a competitive subleading instability.
In combination with electron-phonon interactions~\cite{Brydon2014, Wu2017}, it is possible that this subleading $p$-wave state becomes leading.
A symmetry-breaking strain field could have a similar effect, but only if it is sufficiently large.

Despite their unusual superconductivity, neither \ce{SnTe} nor \ce{Bi2Se3} have strong local electronic correlations or nearly critical collective modes, which was one of the motivations for the current work:
Is parity-mixing and spin-orbit coupling enough to obtain unconventional superconductivity, even in mundane weakly correlated systems?
And can such a mechanism deliver unconventional pairing as the leading instability?
The proposed mechanism answers both in the affirmative.

\begin{acknowledgments}
We thank Avraham Klein, Rafael M.\ Fernandes, Erez Berg, and Jonathan Ruhman for useful discussions.
This work was supported by the Deutsche Forschungsgemeinschaft (DFG, German Research Foundation) -- TRR 288-422213477 Elasto-Q-Mat, project B01.
\end{acknowledgments}

\appendix

\begin{widetext}

\section{Evaluation of the polarization bubble} \label{sec:polarization}
The polarization is defined with the convention $\Pi(q) = \mathscr{V}^{-1}(q) - V^{-1}(q)$, where $\mathscr{V}(q_1) \Kd_{q_1 + q_2} = \ev{\phi(q_1) \phi(q_2)}$ is the dressed plasmon propagator.
To lowest order in $N$, it is given by the fermionic bubble diagram [Fig.~\ref{fig:1-loop-diagrams}(a)]
\begin{align}
\Pi(q) &= - N \int \frac{\dd[4]{k}}{(2\pi)^4} \tr G(k) A(k, k+q) G(k+q) A(k+q, k),
\end{align}
where the thermodynamic and $T = 0$ limits were taken,
\begin{align}
G(k) &= \frac{m \one + \iu \mleft[(\omega_k - \iu \mu) \gamma_0 + v (k_x \gamma_1 + k_y \gamma_2)\mright]}{m^2 + (\omega_k - \iu \mu)^2 + v^2 \vb{k}_{\perp}^2} \equiv \frac{B_k}{C_k}
\end{align}
is the bare fermionic Green's function [Eq.~\eqref{eq:G-1-psi}], and
\begin{align}
A(k, p) &= e \gamma_0 + \iu \eta_z (k_z - p_z) \gamma_0 \gamma_3
\end{align}
is the bare vertex [Eq.~\eqref{eq:A-psi}].
For reasons discussed in Sec.~\ref{sec:Dirac-RG}, we only consider the case $v_z = \eta_{\perp} = 0$ [Eq.~\eqref{eq:quasi-2D-param}].

\begin{figure*}
\includegraphics[width=0.85\columnwidth]{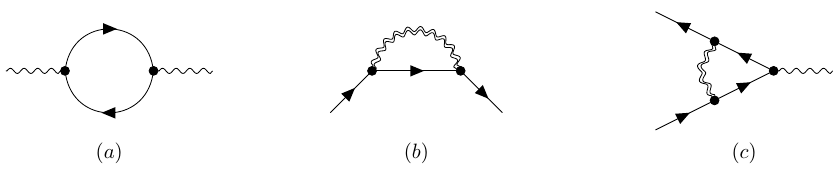}
\caption{The leading contributions to the polarization (a), electronic self-energy (b), and electron-plasmon vertex (c). In the large-$N$ limit, the plasmon propagators needs to be dressed, as indicated by the double wiggly lines. In all cases, the vertices contain both monopole and dipole contributions, as specified in Fig.~\ref{fig:plasmon-fermion-vertex}.}
\label{fig:1-loop-diagrams}
\end{figure*}

First, we consider the retarded real-time polarization $\Pi^R(\nu_q, \vb{q}_{\perp} = \vb{0}, q_z)$ for finite and positive $\mu \geq m$, finite real-time frequencies $\nu_q \neq 0$, arbitrary $q_z$, and vanishing $\vb{q}_{\perp} = (q_x, q_y) = \vb{0}$.
Because the dispersion does not depend on $q_z$, it is straightforward to evaluate the frequency integral to get:
\begin{align}
\Pi(\omega_q, \vb{q}_{\perp} = \vb{0}, q_z) &= \frac{N \Lambda_z}{\pi} \int_{k_F}^{\infty} \frac{k_{\perp} \dd{k_{\perp}}}{2 \pi} \frac{2 m^2 \eta_z^2 q_z^2}{m^2 + v^2 k_{\perp}^2} \mleft(\frac{1}{2 \sqrt{m^2 + v^2 k_{\perp}^2} + \iu \omega_q} + \frac{1}{2 \sqrt{m^2 + v^2 k_{\perp}^2} - \iu \omega_q}\mright),
\end{align}
where $\Lambda_z$ is the $q_z$ cutoff, $q_z \in [- \Lambda_z, \Lambda_z]$, and $k_F$ is the Fermi wavevector, $\mu = \sqrt{m^2 + v^2 k_F^2}$.
The retarded real-time polarization is obtained through the substitution $\iu \omega_q \to \hbar \nu_q + \iu 0^+$.
After applying the Sokhotski-Plemelj formula and evaluating the momentum integral, one obtains the result
\begin{align}
\Pi^R(\nu_q, \vb{q}_{\perp} = \vb{0}, q_z) &= N \Lambda_z \frac{m^2 \eta_z^2 q_z^2}{\pi^2 v^2 \hbar \nu_q} \mleft[\log\abs{\frac{2 \mu + \hbar \nu_q}{2 \mu - \hbar \nu_q}} + \iu \pi \Theta\mleft(\hbar \abs{\nu_q} - 2 \mu\mright)\mright]
\end{align}
which was provided in Eq.~\eqref{eq:Pi-R-quasi2D} of the main text.

For the next two cases, we express the denominator with the help of the Feynman parameterization:
\begin{align}
\frac{1}{C_k C_{k+q}} &= \int_0^1 \dd{x} \frac{1}{\mleft[(1-x) C_k + x C_{k+q}\mright]^2} = \int_0^1 \dd{x} \frac{1}{\mleft[C_p + x (1-x) \mleft(\omega_q^2 + v^2 \vb{q}_{\perp}^2\mright)\mright]^2},
\end{align}
where $p = k + x \, q$.
In the momentum integral we then switch from $k$ to $p$.
Up to terms which are odd in any component of $p$ and thus vanish under the integral, the numerator trace equals
\begin{align}
- \tr B_k A(k, k+q) B_{k+q} A(k+q, k) &= T_1 + T_2 \cdot (1-2x) + T_3 \cdot x(1-x) + T_4 \cdot (v^2 \vb{p}_{\perp}^2 - \omega_p^2),
\end{align}
where
\begin{align}
T_1 &= - 4 (e^2 - \eta_z^2 q_z^2) m^2 - 4 (e^2 + \eta_z^2 q_z^2) \mu^2, &
T_2 &= - 4 (e^2 + \eta_z^2 q_z^2) \mu \iu \omega_q, \\
T_3 &= 4 (e^2 + \eta_z^2 q_z^2) (v^2 \vb{q}_{\perp}^2 - \omega_q^2), &
T_4 &= - 4 (e^2 + \eta_z^2 q_z^2).
\end{align}

When $\mu = 0$, there is an $\SO(3)$ symmetry in the $(\omega_p, v \vb{p}_{\perp})$ variables because of which in the numerator $\vb{p}_{\perp}^2 \to \frac{2}{3} p^2$ and $\omega_p^2 \to \frac{1}{3} v^2 p^2$.
The radial integral is then readily evaluated using dimensional regularization:
\begin{align}
\int_0^{\infty} \frac{p^{2+\epsilon} \dd{p}}{(\Delta^2 + v^2 p^2)^2} &= \frac{(1+\epsilon) \pi}{4 v^4 \cos\frac{\epsilon \pi}{2}} \mleft(\Delta / v\mright)^{\epsilon-1} = \begin{cases}
\displaystyle \frac{\pi}{4 v^3 \Delta}, & \text{for $\epsilon = 0$,} \\[6pt]
\displaystyle - \frac{3 \pi \Delta}{4 v^5}, & \text{for $\epsilon \to 2$.}
\end{cases}
\end{align}
The $\epsilon = 2$ case, which arises during the evaluation of the $T_4$ term contribution, formally diverges.
This divergence is actually spurious.
If instead of radially integrating in frequency and momentum, one first executes the frequency integral and then the momentum integral, one finds a convergent result for the $T_4$ contribution which agrees with the dimensionally regularized result.
The $x$ integrals can be evaluated through a $x \to y = 4 x (1-x)$ substitution with the help of
\begin{align}
\int_0^1 \frac{\dd{y}}{\sqrt{1-y}} \frac{1}{\sqrt{1 + Q^2 y}} &= \frac{2}{Q} \arcctg\frac{1}{Q}, \\
\int_0^1 \frac{y \dd{y}}{\sqrt{1-y}} \frac{1}{\sqrt{1 + Q^2 y}} &= \frac{1}{Q} \mleft[\mleft(1 - \frac{1}{Q^2}\mright) \arcctg\frac{1}{Q} + \frac{1}{Q}\mright], \\
\int_0^1 \frac{\dd{y}}{\sqrt{1-y}} \sqrt{1 + Q^2 y} &= Q \mleft[\mleft(1 + \frac{1}{Q^2}\mright) \arcctg\frac{1}{Q} + \frac{1}{Q}\mright].
\end{align}
The final result, provided in Eq.~\eqref{eq:mu0-polarization} of the main text, is
\begin{align}
\mleft.\Pi(q)\mright|_{\mu = 0} &= N \Lambda_z \frac{\vb{q}_{\perp}^2 (e^2 + \eta_z^2 q_z^2)}{4 \pi^2 \sqrt{\omega_q^2 + v^2 \vb{q}_{\perp}^2}} \mleft[(1-r_q^2) \arcctg r_q + r_q\mright] + N \Lambda_z \frac{2 m^2 \eta_z^2 q_z^2}{\pi^2 v^2 \sqrt{\omega_q^2 + v^2 \vb{q}_{\perp}^2}} \arcctg r_q,
\end{align}
where $q = (\omega_q, \vb{q}_{\perp}, q_z)$, $\vb{q}_{\perp} = (q_x, q_y)$, and $r_q \defeq 2 m / \sqrt{\omega_q^2 + v^2 \vb{q}_{\perp}^2}$.
This $\mu = 0$ polarization reproduces the polarization of Ref.~\cite{DTSon2007} in the $m \to 0$, $\eta_z \to 0$ limit.

When $\omega_q = 0$, but $\mu \geq m$ is finite and positive, we proceed by first evaluating the frequency integral.
We write:
\begin{align}
\Pi(\omega_q = 0, \vb{q}) &= \frac{N \Lambda_z}{\pi} \int_0^1 \dd{x} \int_0^{\infty} \frac{p_{\perp} \dd{p_{\perp}}}{2 \pi} \int_{-\infty}^{\infty} \frac{\dd{\omega_p}}{2 \pi} \mleft(\frac{4 (e^2 + \eta_z^2 q_z^2)}{C_p + x (1-x) v^2 \vb{q}_{\perp}^2} + \frac{- 8 e^2 m^2 - 8 (e^2 + \eta_z^2 q_z^2) v^2 \vb{p}_{\perp}^2}{\mleft[C_p + x (1-x) v^2 \vb{q}_{\perp}^2\mright]^2}\mright).
\end{align}
Note that during the evaluation of the contour integral, one must not overlook the additional Dirac delta function appearing in the second term:
\begin{align}
\int_{-\infty}^{\infty} \frac{\dd{\omega}}{2 \pi} \frac{1}{\Delta + (\omega - \iu \mu)^2} &= \frac{1}{2 \sqrt{\Delta}} \Theta(\Delta - \mu^2), \\
\int_{-\infty}^{\infty} \frac{\dd{\omega}}{2 \pi} \frac{1}{\mleft[\Delta + (\omega - \iu \mu)^2\mright]^2} &= \frac{1}{2 \sqrt{\Delta}} \mleft(\frac{\Theta(\Delta - \mu^2)}{2 \Delta} - \Dd(\Delta - \mu^2)\mright).
\end{align}
The $p_{\perp}$ and $x$ integrals are now readily evaluated.
For $q_{\perp} \leq 2 k_F$, $p_{\perp}$ goes from $\sqrt{k_F^2 - x(1-x) q_{\perp}^2}$ to infinity for all $x$.
For $q_{\perp} > 2 k_F$, one has to separately consider $\abs{x}$ which are smaller and larger than $\frac{1}{2} (1 - \sqrt{1 - 4 k_F^2 / q_{\perp}^2})$.
After some lengthy algebra, one finds that
\begin{align}
\Pi(\omega_q = 0, \vb{q}) &= N g_F e^2 \begin{cases}
\displaystyle 1 + \frac{\eta_z^2 q_z^2}{e^2}, & \text{for $q_{\perp} \leq 2 k_F$,} \\[10pt]
\displaystyle \begin{aligned}
&\mleft(1 + \frac{\eta_z^2 q_z^2}{e^2}\mright) \mleft(1 - \frac{\sqrt{q_{\perp}^2 - 4 k_F^2}}{2 q_{\perp}}\mright) \\[6pt]
&\hspace{10pt} + \frac{\mleft(1 + {\eta_z^2 q_z^2}/{e^2}\mright) v^2 q_{\perp}^2 - 4 m \mleft(1 - {\eta_z^2 q_z^2}/{e^2}\mright)}{4 \mu v q_{\perp}} \arctan\frac{v \sqrt{q_{\perp}^2 - 4 k_F^2}}{2 \mu},
\end{aligned} & \text{for $q_{\perp} > 2 k_F$,}
\end{cases}
\end{align}
where
\begin{align}
g_F &= \frac{\Lambda_z \mu}{\pi^2 v^2}, &
q_{\perp} &= \sqrt{q_x^2 + q_y^2}, &
\mu &= \sqrt{m^2 + v^2 k_F^2}.
\end{align}
In the $\eta_z \to 0$ limit, $\Pi(\omega_q = 0, \vb{q})$ reduces to the expression derived in Refs.~\cite{Gorbar2002, Pyatkovskiy2008}.
The $q_{\perp} \leq 2 k_F$ result is given in Eq.~\eqref{eq:static-smallq-pol} of the main text.

\section{$1$-loop self-energy and vertex diagrams} \label{sec:RG-diagrams}
As in Sec.~\ref{sec:Dirac-RG}, here we consider the quasi-2D limit $v_z = \eta_{\perp} = 0$ [Eq.~\eqref{eq:quasi-2D-param}] with $\mu = T = 0$ [Eq.~\eqref{eq:muT0-limit}].

The fermionic self-energy is defined as $\Sigma(k) = \mathscr{G}^{-1}(k) - G^{-1}(k)$, where $\ev{\psi_{\alpha_1}(k_1) \bar{\psi}_{\alpha_2}(k_2)} = \mathscr{G}_{\alpha_1 \alpha_2}(k_1) \Kd_{k_1 - k_2}$.
It is given by the Fock term [Fig.~\ref{fig:1-loop-diagrams}(b)]:
\begin{align}
\Sigma(k) &= \int \frac{\dd[4]{q}}{(2\pi)^4} A(k, k+q) G(k+q) A(k+q, k) \cdot \mathscr{V}(-q).
\end{align}
For the bare $G(k)$ and $A(k, p)$, see the previous appendix or Sec.~\ref{sec:Dirac-model}.
The Hartree term has been omitted.
Note that the interaction needs to be dressed with the polarization bubble diagram because of the large-$N$ limit.
In a slight abuse of terminology, we shall still call this diagram ``$1$-loop,'' even though a geometric series of loops has been summed up in the interaction.

When $v_z = \eta_{\perp} = \mu = 0$, one finds that:
\begin{align}
A(k, k+q) G(k+q) A(k+q, k) &= \frac{\tilde{T} \cdot \one + T_0 \cdot \gamma_0 + T_1 \cdot \gamma_1 + T_2 \cdot \gamma_2 + T_3 \cdot \gamma_3}{m^2 + (\omega_k + \omega_q)^2 + v^2 (\vb{k}_{\perp} + \vb{q}_{\perp})^2},
\end{align}
where
\begin{align}
\tilde{T} &= m (e^2 - q_z^2 \eta_z^2), &
T_0 &= \iu \mleft(e^2 + q_z^2 \eta_z^2\mright) (\omega_k + \omega_q), \\
T_1 &= - \iu \mleft(e^2 + q_z^2 \eta_z^2\mright) v (k_x + q_x), &
T_2 &= - \iu \mleft(e^2 + q_z^2 \eta_z^2\mright) v (k_y + q_y), &
T_3 &= 2 \iu m e q_z \eta_z.
\end{align}

By expanding in small $k$ and dropping everything odd under $q$, one obtains:
\begin{align}
\mathscr{G}^{-1}(k) &= G^{-1}(k) + \Sigma(k) = Z_m m \one - \iu Z_{\omega} \omega_k \gamma_0 - \iu Z_v v \mleft(k_x \gamma_1 + k_y \gamma_2\mright) + \cdots,
\end{align}
where:
\begin{align}
Z_m &= 1 + \int \frac{\dd[4]{q}}{(2\pi)^4} \frac{e^2 - q_z^2 \eta_z^2}{m^2 + \omega_q^2 + v^2 \vb{q}_{\perp}^2} \cdot \mathscr{V}(-q), \\
Z_{\omega} &= 1 + \int \frac{\dd[4]{q}}{(2\pi)^4} \frac{- (e^2 + q_z^2 \eta_z^2) (m^2 + v^2 \vb{q}_{\perp}^2 - \omega_q^2)}{\mleft[m^2 + \omega_q^2 + v^2 \vb{q}_{\perp}^2\mright]^2} \cdot \mathscr{V}(-q), \\
Z_v &= 1 + \int \frac{\dd[4]{q}}{(2\pi)^4} \frac{(e^2 + q_z^2 \eta_z^2) (m^2 + \omega_q^2)}{\mleft[m^2 + \omega_q^2 + v^2 \vb{q}_{\perp}^2\mright]^2} \cdot \mathscr{V}(-q).
\end{align}

The dressed vertex is defined via
\begin{align}
\ev{\psi_{\alpha_1}(k) \bar{\psi}_{\alpha_2}(p) \phi(q)} - \ev{\psi_{\alpha_1}(k) \bar{\psi}_{\alpha_2}(p)} \ev{\phi(q)} &= \frac{\iu}{\sqrt{L^d \beta}} \mleft[\mathscr{G}(k) \mathscr{A}(k, p) \mathscr{G}(p)\mright]_{\alpha_1 \alpha_2} \mathscr{V}(q) \Kd_{k - p + q}. \label{eq:dressed-vert-def}
\end{align}
Recall that $L^d$ is the volume and $\beta = 1 / (k_B T)$.
To lowest order in $N$, it equals [Fig.~\ref{fig:1-loop-diagrams}(c)]
\begin{align}
\mathscr{A}(k, p) &= A(k, p) - \int \frac{\dd[4]{q}}{(2\pi)^4} A(k, k+q) G(k+q) A(k+q, p+q) G(p+q) A(p+q, p) \cdot \mathscr{V}(-q),
\end{align}
where the interaction again needs to be dressed with the polarization bubble.

Multiplying out the matrices results in gamma matrices of all orders, going from $\one$ and $\gamma_{\mu}$ up to $\gamma_{\mu} \gamma_{\nu} \gamma_{\rho} \gamma_{\sigma} = \LCs_{\mu \nu \rho \sigma} \gamma_0 \gamma_1 \gamma_2 \gamma_3$.
At $k = p = 0$, only the $\propto \gamma_0$ term survives, giving a renormalization of the charge $e$.
At linear order in $k$ and $p$, we find terms $\propto (k_i - p_i) \gamma_0 \gamma_i$ which renormalize $\eta_{\perp}$ and $\eta_z$, but also an additional term $\propto (\omega_k + \omega_p) \one$.
This additional term is irrelevant, just like $\eta_{\perp}$, so we shall neglect it.
All the terms which are higher order in $k$ and $p$ are also irrelevant and thus can be neglected.
After some lengthy algebra, we find that
\begin{align}
\mathscr{A}(k, p) &= Z_e e \gamma_0 + \iu Z_{\eta z} \eta_z (k_z - p_z) \gamma_0 \gamma_3 + \cdots,
\end{align}
where:
\begin{align}
Z_e &= 1 + \int \frac{\dd[4]{q}}{(2\pi)^4} \frac{- (e^2 + q_z^2 \eta_z^2) (m^2 + v^2 \vb{q}_{\perp}^2 - \omega_q^2)}{\mleft[m^2 + \omega_q^2 + v^2 \vb{q}_{\perp}^2\mright]^2} \cdot \mathscr{V}(-q), \\
Z_{\eta z} &= 1 + \int \frac{\dd[4]{q}}{(2\pi)^4} \frac{- (e^2 + q_z^2 \eta_z^2) (- m^2 + v^2 \vb{q}_{\perp}^2 - \omega_q^2)}{\mleft[m^2 + \omega_q^2 + v^2 \vb{q}_{\perp}^2\mright]^2} \cdot \mathscr{V}(-q).
\end{align}
Notice that $Z_e = Z_{\omega}$ and that $Z_{\eta z} = Z_{\omega}$ when $m = 0$.
This is a consequence of exact Ward identities which we prove in the next appendix.

In all the renormalization factors $Z_i$, the frequency and in-plane momentum integrals go up to $\Lambda$, as specified by $\omega_q^2 / v^2 + \vb{q}_{\perp}^2 < \Lambda^2$ [Eq.~\eqref{eq:cutoff-def}].
Differentiating by $\Lambda$ in Eq.~\eqref{eq:RG-flow-expr} thus gives the shell integrals provided in Eqs.~\eqref{eq:RG-flow-final} of the main text.

\section{Ward identities} \label{sec:Ward-id}
Here we prove two Ward identities for the case $v_z = \eta_{\perp} = \mu = 0$ studied in Sec.~\ref{sec:Dirac-RG} and the previous appendix.
Let us start by writing the Euclidean action~\eqref{eq:model-action} in real space:
\begin{align}
\mathcal{S}[\psi, \phi] &= \int_x \bar{\psi} \mleft[m - \gamma_{\mu'} \partial_{\mu'} - \iu \gamma_0 \mleft(e \phi + \eta_z \gamma_3 \partial_3 \phi\mright)\mright] \psi + \frac{1}{2} \epsilon \int_x (\partial_{i} \phi)^2.
\end{align}
Here $x = (\tau, \vb{r})$, $\int_x = \int \dd{\tau} \dd[3]{r}$, and $\partial_{\mu} = \partial / \partial x^{\mu}$.
Summations over repeated $\mu', \nu' \in \{0, 1, 2\}$ indices will be implicit in this appendix, where the prime indicates that $3 = z$ should be excluded.
Temporarily, we have set $v = 1$ and $\epsilon_{\perp} = \epsilon_z = \epsilon$, which we shall later restore.

Under an infinitesimal $\Ugp(1)$ phase rotation $\psi(x) \mapsto \Elr^{\iu \vartheta(x)} \psi(x)$, $\bar{\psi}(x) \mapsto \bar{\psi}(x) \Elr^{- \iu \vartheta(x)}$, $\phi(x) \mapsto \phi(x)$, the action changes by
\begin{align}
\var{\mathcal{S}} &= \iu \int_x \vartheta \, \partial_{\mu'} \mleft(\bar{\psi} \gamma_{\mu'} \psi\mright).
\end{align}
By applying the Schwinger-Dyson equation $\ev{\var{\mathcal{F}}} = \ev{\var{\mathcal{S}} \, \mathcal{F}}$ to the functional $\mathcal{F} = \psi(y) \bar{\psi}(z)$, one obtains the Ward-Takahashi identity:
\begin{align}
(\Kd_{x-y} - \Kd_{x-z}) \ev{\psi(y) \bar{\psi}(z)} &= \ev{\partial_{\mu'} \mleft[\bar{\psi}(x) \gamma_{\mu'} \psi(x)\mright] \psi(y) \bar{\psi}(z)}.
\end{align}
Physically, this identity expresses the conservation of charge within a four-point thermal average.
In Fourier space, it takes the form:
\begin{align}
\ev{\psi(k_1+q) \bar{\psi}(k_2)} - \ev{\psi(k_1) \bar{\psi}(k_2-q)} &= \sum_p \iu q_{\mu'} \ev{\bar{\psi}(p) \gamma_{\mu'} \psi(p+q) \cdot \psi(k_1) \bar{\psi}(k_2)}. \label{eq:WTI1}
\end{align}

Motivated by the above expression, let us introduce for an arbitrary $4 \times 4$ matrix $\Gamma$ the amputated matrix-fermion vertex:
\begin{align}
\mathscr{W}_{\Gamma}(k, q) &\defeq \sum_p \ev{\bar{\psi}(p) \Gamma \psi(p+q) \cdot \mathscr{G}^{-1}(k) \psi(k) \bar{\psi}(k+q) \mathscr{G}^{-1}(k+q)}.
\end{align}

The Ward-Takahashi identity~\eqref{eq:WTI1}, with $k_1 = k$ and $k_2 = k + q$, can now be recast into
\begin{align}
\mathscr{G}^{-1}(k+q) - \mathscr{G}^{-1}(k) &= - \iu \omega_q \mathscr{W}_{\gamma_0}(k, q) - \iu v q_x \mathscr{W}_{\gamma_1}(k, q) - \iu v q_y \mathscr{W}_{\gamma_2}(k, q),
\end{align}
where we have restored $v$.
In particular, this means that:
\begin{align}
\mathscr{W}_{\gamma_{0}}(k, q=0) &= \iu \pdv{}{\omega_k} \mathscr{G}^{-1}(k), \\
\mathscr{W}_{\gamma_{1}}(k, q=0) &= \frac{\iu}{v} \pdv{}{k_x} \mathscr{G}^{-1}(k), \\
\mathscr{W}_{\gamma_{2}}(k, q=0) &= \frac{\iu}{v} \pdv{}{k_y} \mathscr{G}^{-1}(k).
\end{align}
Thus if for small $k$
\begin{align}
\mathscr{G}^{-1}(k) &= Z_m m \one - \iu Z_{\omega} \omega_k \gamma_0 - \iu Z_v v \mleft(k_x \gamma_1 + k_y \gamma_2\mright) + \cdots,
\end{align}
it follows that
\begin{align}
\mathscr{W}_{\gamma_{0}}(k, q=0) &= Z_{\omega} \gamma_0, \label{eq:Wgam0-expr} \\
\mathscr{W}_{\gamma_{1}}(k, q=0) &= Z_{v} \gamma_1, \\
\mathscr{W}_{\gamma_{2}}(k, q=0) &= Z_{v} \gamma_2.
\end{align}

The Schwinger-Dyson equation that follows from varying $\phi(-q)$ with $\mathcal{F} = \psi(k_1) \bar{\psi}(k_2)$ is:
\begin{align}
V^{-1}(q) \ev{\phi(q) \psi(k_1) \bar{\psi}(k_2)} &= \frac{\iu}{\sqrt{\beta L^d}} \sum_p \ev{\bar{\psi}(p) \gamma_0 (e - \iu \eta_z q_z \gamma_3) \psi(p+q) \cdot \psi(k_1) \bar{\psi}(k_2)}.
\end{align}
After using Eq.~\eqref{eq:dressed-vert-def} on the left-hand side, the above becomes:
\begin{align}
\mathscr{A}(k, k+q) &= V(q) \mathscr{V}^{-1}(q) \cdot \mleft[e \mathscr{W}_{\gamma_0}(k, q) - \iu \eta_{z} q_z \mathscr{W}_{\gamma_0 \gamma_3}(k, q)\mright]. \label{eq:SDeq-phi}
\end{align}
If we now assume that for small four-momenta $V(q) \mathscr{V}^{-1}(q) = Z_{\epsilon} + \cdots$ and
\begin{align}
\mathscr{A}(k, k+q) &= Z_e e \gamma_0 - \iu Z_{\eta z} \eta_z q_z \gamma_0 \gamma_3 + \cdots,
\end{align}
as well as exploit Eq.~\eqref{eq:Wgam0-expr}, we obtain the Ward identity $Z_e = Z_{\epsilon} Z_{\omega}$.
In Appendix~\ref{sec:polarization}, we found that $\Pi(q) = \mathscr{V}^{-1}(q) - V^{-1}(q)$ is non-analytic at $q = 0$, which implies that $\Pi(q)$ cannot be Taylor expanded at $q = 0$.
Moreover, there is no canonical decomposition of $\Pi(q)$ into a non-analytic part and analytic part (which could then be expanded around $q = 0$).
Hence no part of $\Pi(q)$ contributes to the renormalization of the bare plasmon propagator.
Consequently, $Z_{\epsilon} = 1$ and we obtain the Ward identity
\begin{align}
Z_e = Z_{\omega}.
\end{align}
Physically, this identity expresses the fact that charge does not renormalize, as we explicitly saw on the $1$-loop level in Appendix~\ref{sec:RG-diagrams}.

Apart from the $\Ugp(1)$ phase rotation symmetry which is associated with charge conservation, in the massless limit there is an additional $\Ugp(1)$ chiral rotation symmetry of the form $\psi(x) \mapsto \Elr^{\iu \vartheta(x) \gamma_3} \psi(x)$, $\bar{\psi}(x) \mapsto \bar{\psi}(x) \Elr^{\iu \vartheta(x) \gamma_3}$, $\phi(x) \mapsto \phi(x)$.
Analogous manipulations to the previous give the Ward-Takahashi identity
\begin{align}
\gamma_3 \mathscr{G}^{-1}(k+q) + \mathscr{G}^{-1}(k) \gamma_3 &= \iu \omega_q \mathscr{W}_{\gamma_0 \gamma_3}(k, q) + \iu v q_x \mathscr{W}_{\gamma_1 \gamma_3}(k, q) + \iu v q_y \mathscr{W}_{\gamma_2 \gamma_3}(k, q),
\end{align}
which implies
\begin{align}
\mathscr{W}_{\gamma_0 \gamma_3}(k, q=0) &= Z_{\omega} \gamma_0 \gamma_3, \\
\mathscr{W}_{\gamma_1 \gamma_3}(k, q=0) &= Z_{v} \gamma_1 \gamma_3, \\
\mathscr{W}_{\gamma_2 \gamma_3}(k, q=0) &= Z_{v} \gamma_2 \gamma_3.
\end{align}
From Eq.~\eqref{eq:SDeq-phi} we now obtain the Ward identity
\begin{align}
Z_{\eta z} = Z_{\omega},
\end{align}
where we used the fact that $\mathscr{W}_{\gamma_0}(k, q)$ cannot be linear in $q_z$ because of horizontal reflection symmetry.
In the massless limit, the chiral symmetry thus protects the out-of-plane electric dipole moment $\eta_z$ from renormalizing.

\end{widetext}

\bibliography{dipole-SC-references}

\end{document}